\documentclass[fleqn,usenatbib]{mnras}

\usepackage{newtxtext,newtxmath}
\usepackage[T1]{fontenc}
\usepackage{ae,aecompl}

\usepackage{lineno}
\modulolinenumbers[5]

\usepackage{hyperref}
\usepackage{amsmath}
\usepackage{amssymb}

\usepackage{etoolbox}
\makeatletter
\patchcmd\@combinedblfloats{\box\@outputbox}{\unvbox\@outputbox}{}{\errmessage{\noexpand patch failed}}
\makeatother

\usepackage{psfrag} % for nicer fonts in figures

\usepackage{media9} % For including the animation

\usepackage[utf8]{inputenc} % Allows to write utf8 encoding directly
\usepackage{graphicx}

\usepackage{relsize}%for a bigger sum / smaller symbols
\usepackage{pbox}
\usepackage[usenames,dvipsnames]{color}
\definecolor{aliceblue}{rgb}{0.94, 0.97, 1.0}
\definecolor{lightgray}{rgb}{0.83, 0.83, 0.83}
\usepackage{bm}% for fat greek letters!

\usepackage{array}
\usepackage{ulem}

\newcommand{\jcop}{J. Comput. Phys.}
\newcommand{\vect}[1]{\boldsymbol{#1}}

\title[Striped jets from small scale magnetic field]{Striped Blandford/Znajek jets from advection of small scale magnetic field}

\author[J. F. Mahlmann, A. Levinson and M. A. Aloy]
{J. F. Mahlmann$^1$\thanks{jens.mahlmann@uv.es}, A. Levinson$^2$, and M. A. Aloy$^1$\thanks{miguel.a.aloy@uv.es}\\ $^1$Departament d'Astronomia i Astrofísica, Universitat de València, 46100, Burjassot, Spain\\
 $^2$The Raymond and Beverly Sackler School of Physics and Astronomy, Tel Aviv University, Tel Aviv 69978, Israel}

\date{\today. Accepted 2020 March 31. Received 2020 March 2; in original form 2020 January 9}

\pagerange{\pageref{firstpage}--\pageref{lastpage}} \pubyear{2019}

\begin{document}
	
	\maketitle
	
	\label{firstpage}
	
	\begin{abstract}
		Black hole - accretion disc systems are the central engines of relativistic jets from stellar to galactic scales. We numerically quantify the unsteady outgoing Poynting flux through the horizon of a rapidly spinning black hole endowed with a rotating accretion disc. The disc supports small-scale, concentric, flux tubes with zero net magnetic flux. Our General Relativistic force-free electrodynamics simulations follow the accretion onto the black hole over several hundred dynamical timescales in 3D. For the case of counter-rotating accretion discs, the average process efficiency reaches up to $\left\langle\epsilon\right\rangle\approx 0.43$, compared to a stationary energy extraction by the Blandford/Znajek process. The process efficiency depends on the cross-sectional area of the loops, i.e. on the product $l\times h$, where $l$ is the radial loop thickness and $h$ its vertical scale height. We identify a strong correlation between efficient electromagnetic energy extraction and the quasi-stationary setting of ideal conditions for the operation of the Blandford/Znajek process (e.g. optimal field line angular velocity and fulfillment of the so-called Znajek condition). Remarkably, the energy extraction operates intermittently (alternating episodes of high and low efficiency) without imposing any large-scale magnetic field embedding the central object. Scaling our results to supermassive black holes, we estimate that the typical variability timescale of the system is of the order of days to months. Such timescales may account for the longest variability scales of TeV emission observed, e.g. in M87.
	\end{abstract} 
	
	
%	\linenumbers
	
	%--------------------------------------------------------------------
%--------------------------------------------------------------------
\section{Introduction}
\label{sec:introduction}
%--------------------------------------------------------------------
%--------------------------------------------------------------------

The recent observations of the \textit{shadow} around the event horizon of the black hole (BH) in the nucleus of the elliptical galaxy M87 \citep{ETH_2019a} have sparked further interest for numerical models of BHs surrounded by magnetised accretion discs (ADs). 
Strong magnetic fields in these astrophysical systems (BH/AD) have been established as efficient mediators to tap a fraction of the gravito-rotational energy of the system and power relativistic jets \citep[e.g.][and references therein]{Blandford+2019,Marti2019} from stellar (e.g. jets associated with microquasars and $\gamma$-ray bursts; GRBs) to galactic scales (e.g. kiloparsec-scale jets associated with active galactic nuclei; AGNs). Two basic mechanisms of energy extraction out of BH/AD systems have been extensively studied, namely the Blandford/Znajek (BZ) mechanism \citep{Blandford1977} and the magneto-centrifugal (MC) jet launching \citep{Bisnovatyi-Kogan_1976,Blandford1982}. The BZ process posits the existence of large-scale poloidal magnetic fields threading the horizon of a spinning BH, which carry away the reducible energy of the central object to infinity. In contrast, the MC mechanism needs a large-scale magnetic field anchored to the AD (with an adequate inclination away from the vertical) in order to provide sufficient magneto-centrifugal thrust to generate a plasma outflow. 

Mounting evidence for the important role of magnetic fields in the jet launching process has been provided by General Relativistic magnetohydrodynamics (GRMHD) simulations \citep[e.g.][]{Koide2000,Hawley2006,McKinney2004,Komissarov2009a,Tchekhovskoy2011,McKinney2012,Tchekhovskoy2012,Tchekhovskoy2012a,McKinney+2014,Skadowski+2016,Chatterjee+2019,Liska+2019,Vourellis+2019}. In many of these numerical models \citep[e.g.][]{Koide2000,McKinney2004,Hawley2006, Komissarov2009a, McKinney2012,Tchekhovskoy2011,Tchekhovskoy2012,Tchekhovskoy2012a}, the initial magnetic structure is not self-consistently generated, but assumed to have some large-scale poloidal topology as a starting point. Because of the numerical challenge that simulating a thin disc represents, most GRMHD simulations begin with a geometrically thick torus, even though thin discs may be more physically suited, e.g., for BH/AD systems in AGNs. Furthermore, the 
micro-physics associated with electron and ion heating and cooling, energy transfer between electrons 
and ions, and plasma production in the force-free section of the magnetosphere is poorly understood. They are either ignored or prescribed using ad hoc assumptions.

While advection of large scale magnetic fields by the hot accretion flow can lead to the efficient production of a powerful jet \citep[particularly in the Magnetically Arrested Disc - MAD - regime][]{Narayan_2003PASJ...55L..69}, as demonstrated by GRMHD simulations \citep[e.g.][]{Igumenshev_2008ApJ...677..317}, 
some key issues remain unresolved. The most burning one is the dissipation of the jet magnetic field. 
The current-driven kink instability has been considered as a potential 
mechanism to generate strong distortions that can ultimately lead to the formation of current sheets and turbulence, however, under which conditions this instability develops, and at what scales, is yet unclear. 
Recent numerical experiments of the relativistic kink	instability of Poynting dominated jets \citep{Bromberg_2019ApJ...884...39, Davelaar_2019arXiv191013370} invoke special (and artificial) setups for numerical convenience  \citep[e.g., non-rotating, stationary cylindrical flux tubes, but see][for inclusion of rotation]{Mizuno_2012ApJ...757...16}. These experiments indicate that cylindrical jets 
with dominant toroidal fields should become kink unstable, and may be disrupted over a timescale of $\sim 10^2 R/c$ or longer, where $R$ is the cross-sectional radius of the jet. Favorable conditions for such dissipation are anticipated
in strong collimation sites \citep{Bromberg2016}. Such strong collimation is expected in GRBs during the propagation of the jet inside the star \citep{Aloy_2000ApJ...531L.119, Bromberg2016, Obergaulinger_2017MNRAS.469L..43, Aloy_2018MNRAS.478.3576}, and is occasionally seen in AGN reconfinement zones, e.g., the HST-1 knot in M87. However, these zones are usually located far from the BH, typically at radii $10^5 -10^7 r_g$.
Yet, in many objects dissipation is seen or inferred on much smaller scales. Velocity maps of the inner M87
jet \citep{Mertens_2016} and its apparent limb brightening \citep{Kim2018} are indicative of a dissipative 
boundary layer (or sheath) down to horizon scales, as also suggested by the recent ETH analysis \citep{EHT2019v}.
The nature of the dissipation mechanism in the sheath is unclear, but it is, most likely, unrelated to the kink instability \citep[the sheath may result from the turbulence induced by the non-linear development of Kelvin-Helmholtz instabilities, see e.g.,][]{DeYoung_1993ApJ...405L..13, Aloy_2000ApJ...528L..85, Aloy_2008ApJ...681...84}. 

The rapid variability observed in many blazars and other radio loud AGNs, 
and in particular the extreme gamma-ray flares, also 
require rapid dissipation of the inner jet, on scales at which the jet is not expected to be prone to instabilities. 
Notable examples are M87 \citep{Acciari2009,Aharonian2003} and IC310 \citep{Aleksic2014}
that exhibit occasional strong flares with durations as short as one day (roughly $r_g/c$) in M87, 
and a few minutes in IC310 \citep{Aleksic2014}. These flares are likely produced in the innermost regions,
close to the BH (but see \citealt{Barkov2012} for a different interpretation). It has been proposed that the 
variable TeV emission in M87 (and conceivably IC310) may originate from a magnetospheric spark gap located at the base of a jet \citep{Levinson2000,Neronov2007,Levinson2011,Hirotani2016,Hirotani2016a,Hirotani2017,Levinson2017,Lin2017}.
Recent attempts to study this process using 1D GRPIC simulations \citep{Levinson2018,Chen2019} confirm that such
gaps are self-sustained when the pair production opacity contributed by the disc emission is large enough, and
that they are potential sources of intermittent TeV emission. However, these simulations are local and, thus, missing 
information about the feedback of the global magnetosphere. 
Global 2D GRPIC simulations, as those described in \cite{Parfrey_2019PhRvL}, may be able to shed more light on 
the gap emission. 

Alternatively, dissipation and rapid variability can more naturally arise from advection of
small scale magnetic fields by the accretion flow, as demonstrated by recent 2D General Relativistic \citep{Parfrey2015} and 3D special relativistic \citep{Yuan2019,Yuan2019a} force-free 
simulations. Along this idea, \cite{Giannios2019} have argued that the variability  timescales are related to the growth of the magnetorotational dynamo in the AD. But can accretion of small scale magnetic field lead
to formation of a striped relativistic jet with a substantial mean power? This is 
the prime question addressed in this paper.

The picture envisaged here is inspired by the model described in \citet{Uzdensky2008}. 
They describe the AD corona as a statistical ensemble of magnetic loops \citep{Coroniti1985,Tout1992,Hughes2003}, continuously emerging from and submerging into the disc due to magnetic buoyancy (or a \textit{boiling magnetic foam}). Reconnection between these loops is able to rapidly dissipate magnetic energy \citep{DiMatteo1999} and to produce spatially extended (loop) structures in the AD and its corona \citep{Romanova1998,Uzdensky2008}. The existence of such loop structures of zero net flux in the AD was acknowledged by \citet{McKinney2005} as possibly relevant to power BZ-like energy flows in thin BH/AD systems. 
In the  \textit{coronal mechanism} \citep{Beckwith2009}, the magnetic flux is advected as a consequence of the reconnection of loops across the equator, which induce the formation of magnetic loops in the corona. The poloidal magnetic fields added to the accretion funnel in this way are a requirement for the formation of a BZ-type jet. 
The premise of accreting (zero net flux) loops was recently used in 2D axisymmetric, 
General Relativistic force-free electrodynamics (GRFFE) simulations by \citet{Parfrey2015} to confirm an efficient working of the BZ process. However, whether a similar behaviour is expected in 3D is yet an open issue. For instance, \citet{Beckwith2008} find a significant sensitivity of the jet power on the topology of the accreted (small net flux) magnetic field. In this paper we present results of 3D GRFFE simulations of loop accretion, using a similar setup to that invoked in \cite{Parfrey2015}. We find that substantial power can be extracted in the form of a striped BZ jet for a range of conditions, and that dissipation in current sheets at the jet boundary is anticipated due to interaction of loops.

This work is organised as follows. In section~\ref{sec:magnetosphere} we introduce the notation to deal with the General Relativistic problem at hand, which includes a Kerr BH (sec.\,\ref{sec:kerr_solution}) surrounded by an \textit{idealised} AD, where loops of alternate polarity and zero net magnetic flux are set up (\ref{sec:loop_systems}). 
We provide the equations of GRFFE (section~\ref{sec:force-free}) implemented for simulations conducted on the infrastructure of the \texttt{Einstein Toolkit} (supplemented by appendix~\ref{sec:augmented_system}). Section~\ref{sec:simulations} summarises numerical simulations of accreting tubes of magnetic flux in magnetospheres of rapidly spinning BHs (with a dimensionless rotation rapidity $a^*=0.9$) for both counter-rotating and co-rotating disc systems. 
Our work improves on the 2D axisymmetric model of \citet{Parfrey2015} by considering full-fledged 3D BH/AD systems. Inclusion of 3D is insurmountable to properly understand the dynamics of the electromagnetic fields developed in the BH magnetosphere, where finite resistivity (in our models of numerical origin) may yield episodes of fast dissipation and, thus, rapid variability of the plasma in the vicinity of the BH. Besides, our models aim to explore systematically the dependence of the variability timescales on the (simple) parameterisation of the magnetic loops in the AD.
We discuss general trends through the chosen parameters in section~\ref{sec:discussion} and provide relevant scaling to supermassive BHs. Section~\ref{sec:conclusions} concludes the astrophysical implications of the presented simulations.

	%--------------------------------------------------------------------
	%--------------------------------------------------------------------
	\section{Magnetosphere setup and evolution}
	\label{sec:magnetosphere}
	%--------------------------------------------------------------------
	%--------------------------------------------------------------------
	
	The following sections and the \texttt{Einstein Toolkit} employ units where $M_\odot=G=c=1$, which sets the respective time and length scales to be $1M_\odot\equiv 4.93\times 10^{-6}\text{ s}\equiv 1477.98\text{ m}$. This unit system is a variation of the so-called system of \textit{geometrised units} \citep[as introduced in appendix F of][]{Wald2010}, with the additional normalisation of the mass to $1M_\odot$ \citep[see also][on unit conversion in the \texttt{Einstein Toolkit}]{Mahlmann2019}.
	
	%--------------------------------------------------------------------
	%--------------------------------------------------------------------
	\subsection{The Kerr solution}
	\label{sec:kerr_solution}
	%--------------------------------------------------------------------
	%--------------------------------------------------------------------
	
	The Kerr solution embodies the geometry of
	a spinning BH of mass $M$ and specific angular
	momentum $a=J/M$, where $J$ is the angular momentum and $a^*=J/M^2$ is the dimensionless rotation rapidity \citep[cf.][]{Frolov2011}. In Boyer-Lindquist coordinates, the line element of the Kerr
	metric is
	\begin{align}
	\begin{split}
	\text{d} s^2&=\:-\left(1-\frac{2Mr}{\Sigma}\right)\text{d} t^2-\frac{4Mar\sin^2\theta}{\Sigma}\:\text{d} t\text{d} \phi\\
	&+\frac{\Sigma}{\Delta}\:\text{d} r^2+\Sigma\:\text{d}\theta^2+\frac{A\sin^2\theta}{\Sigma}\:\text{d}\phi^2\,,\label{eq:BoyerLindquist}
	\end{split}
	\end{align}
	\begin{align}
	\begin{split}
	\Sigma:=&\: r^2+a^2\cos^2\theta\,,\\
	A:=&\:\left(r^2+a^2\right)^2-\Delta\: a^2\sin^2\theta\,,\\
	\Delta:= &\:r^2-2 M r+a^2 :=\left(r-r_+\right)\left(r-r_-\right)\,,
	\end{split}
	\end{align}
	where $r_\pm$ represent the locations
	of the inner and outer horizons of the BH,
	respectively:
	\begin{align}
	r_\pm= M\pm\sqrt{M^2-a^2}\,.
	\end{align}
	The BH mass is a scale parameter of the presented line-element (\ref{eq:BoyerLindquist}), i.e. one can write $\text{d} s^2=M^2 \text{d}\tilde{s}^2$ where $\text{d}\tilde{s}^2$ is a (dimensionless) function of $a^*$ only \citep[cf.][]{Frolov2011}. The frame-dragging frequency induced by the rotation of the BH is
	\begin{align}
	\Omega:=\:2aMr/A\,,
	\label{eq:framedragging}
	\end{align}
	which is also the angular velocity of the (local)
	\textit{zero angular momentum observer} or ZAMO \citep[cf.][]{Thorne1986}, i.e.,
	$\Omega=(d\phi/dt)_{\rm ZAMO}$. 
	At the outer event horizon, the frame dragging frequency reads
	\begin{align}
	\Omega_{\rm BH}:=\Omega(r=r_+) = \frac{a}{2Mr_+}\,.\label{eq:frame_dragging}
	\end{align}
	The redshift which accounts for the lapse of proper time $\tau$ in the
	ZAMO frame with respect to the global (Boyer-Linquist) time $t$, thus,
	$\alpha=(d\tau/ dt)_{\rm ZAMO}$ is
	\begin{align}
	\alpha:=\sqrt{\frac{\Sigma\Delta}{A}}\,.\label{eq:lapse}
	\end{align}
	While quantities in Boyer-Lindquist
	coordinates are represented in a spatial basis made by the set of
	orthogonal vectors $\{\partial_i\} = \{\boldsymbol{e}_i\}$, the
	local ZAMO observers have an attached triad
	$\{\boldsymbol{\hat{e}}_i\}=\{\boldsymbol{e}_i / \sqrt{g_{ii}}\}$,
	where the index $i$ runs over the three spatial coordinates
	$(r,\theta,\phi)$. $g_{ii}$ are the diagonal components of the
	metric tensor, namely
	\begin{align}
	g_{rr} = \frac{\Sigma}{\Delta}\,, \qquad g_{\theta\theta} = \Sigma\,, \qquad g_{\phi\phi}=\frac{A\sin^2{\theta}}{\Sigma}\,.
	\end{align}
	The ZAMO's four velocity is $n_\mu=\left(-\alpha,0,0,0\right)$ and may be used to introduce a projection tensor on the spatial components of a suitable $3+1$ decomposition of spacetime:
	\begin{align}
	\gamma_{\mu\nu}=g_{\mu\nu}+n_\mu n_\nu\label{eq:metric_decomposition}
	\end{align}
	The determinants of the metric tensors will be denoted by $g$ or $\gamma$, respectively.
	
	%--------------------------------------------------------------------
	%--------------------------------------------------------------------
	\subsection{Force-free electrodynamics}
	\label{sec:force-free}
	%--------------------------------------------------------------------
	%--------------------------------------------------------------------
	
	In analogy to \citet{Komissarov2004} and \citet{Parfrey2017}, we solve Maxwell's equations in the force-free limit:
	\begin{align}
	\nabla_\nu F^{\mu\nu}=&\:I^\mu\label{eq:MaxwellCovariantI}\\
	\nabla_\nu\hspace{2pt} ^{\ast}\hspace{-2pt}F^{\mu\nu}=&\: 0\label{eq:MaxwellCovariantII}
	\end{align}
	Here, $F^{\mu\nu}$ and $^{\ast}F^{\mu\nu}$ are the Maxwell tensor and
	its dual, respectively. $I^\mu$ is the electric current four
	vector associated to the charge density $\rho=\alpha I^t$, and the current three vector $J^i=\alpha I^i$. $\nabla$ denotes the covariant derivative, Greek indices reflect arbitrary spacetime quantities, Latin indices will refer to the coordinate directions of a $3+1$ spacetime decomposition (see eq.~\ref{eq:metric_decomposition}). We separately evolve the continuity equation of total electric charge
	\begin{align}
	\nabla_\nu I^\nu=\:0\label{eq:ChargeConservation}\,,
	\end{align}
	in order to ensure conservation of (total) electric charge in the computational domain. \cite{Komissarov2004} introduces the equivalent of the classical field quantities $\mathbf{E}$, $\mathbf{B}$, $\mathbf{D}$, and $\mathbf{H}$ in a $3+1$ decomposition of spacetime:
	\begin{align}
	E_i =&\:F_{it}\label{eq:MacroscopicE} ,\\
	B^i =&\:\frac{1}{2}e^{ijk}F_{jk} ,\\
	D^i =&\:\frac{1}{2}e^{ijk} {^*}\hspace{-2pt}F_{jk}, \\
	H_i =&\:{^*}\hspace{-2pt}F_{ti}\label{eq:MacroscopicH} ,
	\end{align}
	where $e^{ijk}=\alpha\eta^{0ijk}$, with the volume element $\eta^{\mu\nu\lambda\zeta}=[\mu\nu\lambda\zeta]/\sqrt{-g}$, and the completely antisymmetric Levi-Civita symbol $[\mu\nu\lambda\zeta]$. These equivalents to the classical electric field and magnetic induction as well as the electric displacement and the magnetic field encode the geometry of spacetime (i.e. the lapse in time and frame-dragging of space) as non-vacuum effects in the full set of macroscopic Maxwell equations. In order to solve eqs.~(\ref{eq:MaxwellCovariantI}) and~(\ref{eq:MaxwellCovariantII}), one specifies the following constitutive relations \citep[cf.][]{jackson1999}:
	\begin{align}
	\mathbf{E}=&\alpha \mathbf{D}+\boldsymbol{\beta}\times\mathbf{B}\,,\\
	\mathbf{H}=&\alpha \mathbf{B}-\boldsymbol{\beta}\times\mathbf{D}\,.
	\end{align} 
	We may now write the Maxwell tensor as measured by the normal observer (ZAMO) in terms of the macroscopic field quantities \citep[cf.][]{Anton2006}:
	\begin{align}
	F^{\mu\nu}&=\: n^\mu D^\nu-D^\mu n^\nu-e^{\mu\nu\lambda\zeta}B_\lambda n_\zeta\label{eq:MaxTensorI} , \\
	{^*}\hspace{-2pt}F^{\mu\nu}&=\: -n^\mu B^\nu+B^\mu n^\nu-e^{\mu\nu\lambda\zeta}D_\lambda n_\zeta , \label{eq:MaxTensorII}
	\end{align}
	To build up a stationary magnetosphere around the central BH, it is necessary to guarantee
	that there are either no forces acting on the system or, more
	generally, that the forces of the system are in equilibrium. The latter condition
	implies that the electric 4-current 
	%$J^{\mu}$ 
	$I^\mu$
	satisfies the force-free condition \citep{Blandford1977}:
	\begin{align}
	F_{\mu\nu}I^\nu=0\,,
	\label{eq:ForceFree}
	\end{align}
	With the above definition (\ref{eq:MaxTensorI}), this condition (\ref{eq:ForceFree}) reduces to
	\begin{align}
	D^\mu B_\mu=0\qquad\Leftrightarrow\qquad{^*}\hspace{-2pt}F_{\mu\nu}F^{\mu\nu}=0\,.
	\label{eq:FFCondICov}
	\end{align}
	The component of the electric field parallel to the magnetic field vanishes. A second condition of magnetic dominance is given by
	\begin{align}
	F_{\mu\nu}F^{\mu\nu}>0\,.
	\label{eq:FFCondIICov}
	\end{align}
	In the language of the full system of Maxwell's equations in $3+1$ decomposition, expressions (\ref{eq:FFCondICov}), and (\ref{eq:FFCondIICov}) respectively read
	\begin{align}
	\mathbf{D}\cdot\mathbf{B}=0\label{eq:FFCondI} \\
	\mathbf{B}^2-\mathbf{D}^2\geq 0\label{eq:FFCondII}\,.
	\end{align}
	Conditions (\ref{eq:FFCondI}) and (\ref{eq:FFCondII}), as well as the conservation condition $\partial_t\left(\vect{D}\cdot\vect{B}\right)=0$ can be combined in order to obtain an explicit expression for the so-called force-free current $I^\mu_{\textsc{ff}}$ \citep[cf.][]{Komissarov2011,Parfrey2017}:
	\begin{align}
	\begin{split}
	I^\mu_\textsc{ff}=&\:\rho n^\mu+\frac{\rho}{\mathbf{B}^2}e^{\nu\mu\alpha\beta}n_\nu D_\alpha B_\beta\\
	&+\frac{B^\mu}{\mathbf{B}^2} \:e^{\alpha\beta\mu\sigma}n_\sigma\left(B_{\mu;\beta}B_\alpha-D_{\mu;\beta}D_\alpha\right)
	\label{eq:FFCurrent}
	\end{split}
	\end{align}
	In practice, the combination of the force-free current (\ref{eq:FFCurrent}) as a source-term to eq.~(\ref{eq:MaxwellCovariantI}) with numerically enforcing conditions (\ref{eq:FFCondI}) and (\ref{eq:FFCondII}) restricts the evolution to the force-free regime. The discussion of techniques in order to ensure a \textit{physical} \citep[cf.][]{McKinney2006} evolution of numerical force-free codes can be found throughout the literature \citep[e.g.][]{Lyutikov2003,Komissarov2004,Palenzuela2010,Alic2012,Paschalidis2013,Carrasco2016,Parfrey2017}. A review of the employed conservative system of equations, and techniques to minimise numerical errors is given in appendix~\ref{sec:augmented_system}, as well as \citet{Mahlmann2019}.
	
	%--------------------------------------------------------------------
	%--------------------------------------------------------------------
	\subsection{Magnetic loop accretion systems}
	\label{sec:loop_systems}
	%--------------------------------------------------------------------
	%--------------------------------------------------------------------
	
	%
	\begin{figure}
		\centering
		\includegraphics[width=0.425\textwidth]{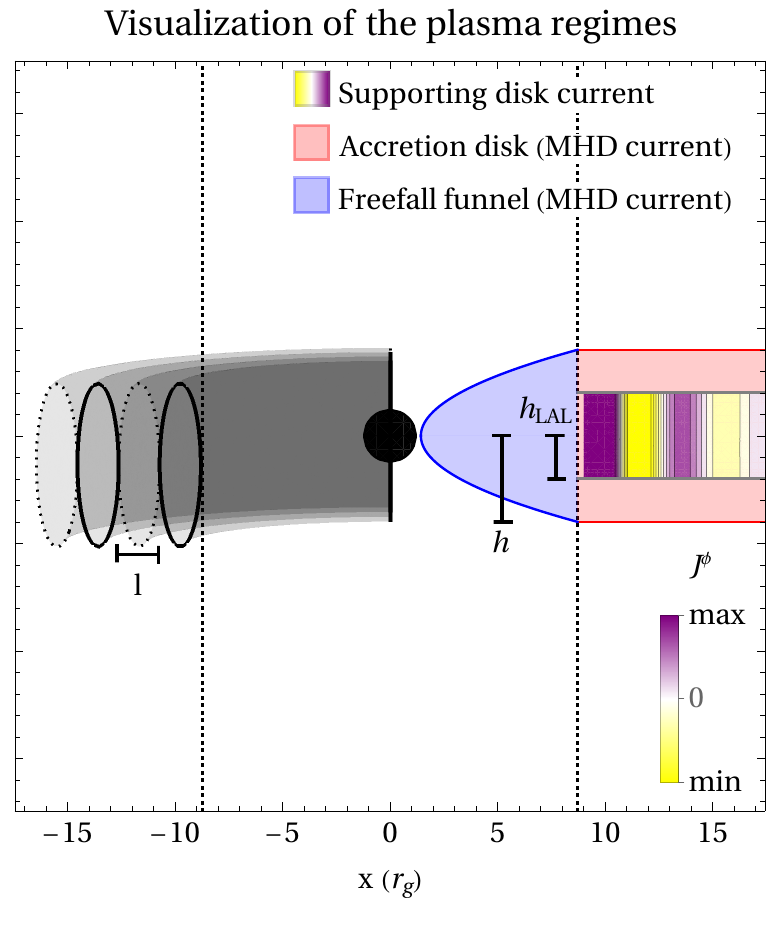}
		\vspace{-9pt}
		\caption{Schematic visualisation of the employed accretion disc model. \textit{Left half:} Sketch of the flux tubes (length $l$ and scale height $h$) supported by the AD model. Flux tubes with opposite polarity are denoted by enclosing their cross-section with dashed and solid-line ellipses.  \textit{Right half:} Structure of the different currents in the simulation domain. The supporting current maintains the loop structure over long periods of time by imposing an alternating (toroidal) current density in the LAL (violet-to-yellow shades). The MHD currents in the UAL (red shade) introduce the accretion dynamics in the system for the \textit{assumed} equatorial velocity field. In the white region we employ full force-free electrodynamics (see section~\ref{sec:force-free}).}
		\label{fig:discsetup}
	\end{figure}

	Since the force-free approximation corresponds to the limit of vanishing plasma inertia, it is not possible to construct an AD self-consistently around a central BH. Thus, we resort to introducing a suitable AD setup, which is a crucial ingredient in our models and resembles, to some extent, the procedure sketched in \citet{Parfrey2015}. We consider the force-free evolution of a system composed by a BH surrounded by an (idealised) equatorial AD of finite vertical extension. There, the magnetic field can be anchored. In practical terms, the \textit{ad-hoc} equatorial structure that we construct serves for the purpose of providing conditions that may mimic the accretion process found in self-consistently built ADs. 
		
	In our model, the equatorial region around the BH is set up using different types of electric currents (Figure~\ref{fig:discsetup}), which serve two purposes. First, they mimic the accretion flow in the AD, and second, they drive the plasma from the innermost stable circular orbit (ISCO) to the BH. The AD is split into two regions located at different vertical distances from the equator. In the innermost region, $|z|\le h_{\textsc{lal}}$, referred to in the following as the \textit{lower accretion layer} (LAL; violet-to-yellow shades in Figure~\ref{fig:discsetup}), we impose a toroidal current (eq.~\ref{eq:InitCurrent}), whose effect is the generation of a set of concentric magnetic flux tubes that move radially inward at a prescribed \textit{accretion speed}. In the limit of GRFFE, there is no actual fluid and, hence, the fluid velocity is not even a variable of the governing equations. Thus, the accretion speed is not the self-consistent result of the physical transport of angular momentum towards the exterior. Additionally, we impose the rotational speed of the AD in our models. Finding a poloidal current that effectively induces the sought for rotational profile is not an easy task. Instead, we specify the rotational profile resorting to a different procedure, namely, adding a further (resistive MHD) layer surrounding the LAL. Extending vertically for $h_{\textsc{lal}}\le |z|\le h$, ($h$ is, thus, the disc half-thickness) we define the \textit{upper accretion layer} (UAL). There, we prescribe a resistive MHD current (red shaded region in Figure~\ref{fig:discsetup}, eq.\,\ref{eq:MHDcurrent}) that replaces the force-free current in the GRFFE equations. In this resistive MHD current, besides the radial motion, the rotational profile of the AD can be easily specified (see eq.~\ref{eq:v_0max} below). We drive the \textit{free-fall} of plasma from the ISCO to the BH by prescribing a free-falling geodesic motion in the equatorial region (Figure~\ref{fig:discsetup}, blue shaded). We refer to the paraboloidally shaped region $r_+<r_c<r_{\rm ISCO}$ and $\left|z\right|\leq h\sqrt{(r_c-r_+)/(r_{\rm ISCO}-r_+)}$ as \textit{plunging region}. Here, $r_c=\sqrt{g_{\phi\phi}}$ is the cylindrical radius from the central object. As in the UAL, the kinematics of the plunging region are prescribed by employing a resistive MHD current whose velocity field follows a free-falling geodesic motion in the equatorial plane, matching the angular velocity to the BH rotation smoothly (see eq.~\ref{eq:frame_dragging}). Since the plunging region is relatively thin, we employ an exclusively radially dependent velocity prescription all over its volume, regardless of the vertical distance to the equator.
	
	In the following paragraphs, we describe the specific form of the currents employed in every region in detail. We begin by specifying the form of the resistive MHD current used in the UAL. For that, we need to provide a closure relation, the Ohm's law \citep[see, e.g.,][]{Baumgarte2003,Palenzuela2010,Parfrey2017}, for which we take
	\begin{align}
	\begin{split}
	I_\mu-\tilde{\rho}u_\mu = \sigma F_{\mu\nu}u^\nu\,.
	\label{eq:OhmsLaw}
	\end{split}
	\end{align}
	Here, $\sigma$ is the plasma conductivity, $\tilde{\rho}=-I^\nu u_\nu$ denotes the charge density as seen by an observer comoving with the fluid \citep{Baumgarte2003}, while $\rho=-I^\nu n_\nu$ (with the definitions given in sections~\ref{sec:force-free} and~\ref{sec:kerr_solution}) is the charge density observed by a normal observer \citep[][]{Komissarov2004}. For the four-velocity $u^\mu$ we define the Lorentz factor
	\begin{align}
	\begin{split}
	W=-n_\mu u^\mu=\alpha u^t\, ,
	\label{eq:4velocity}
	\end{split}
	\end{align}
	and the normal observer 3-velocity \citep[cf.][]{Baumgarte2003}
	\begin{align}
	v^i=\frac{u^i}{u^t}\,.
	\label{eq:3velocity}
	\end{align}
	Inserting the definition of fields from eqs.~(\ref{eq:MacroscopicE}) - (\ref{eq:MacroscopicH}) into eq.~(\ref{eq:OhmsLaw}), and using the notation introduced in eqs.~(\ref{eq:4velocity}) and (\ref{eq:3velocity}), one obtains the resistive MHD current used in the presented simulations:
	\begin{align}
	\begin{split}
	I^i = W\sigma\left[E^i+e^{ijk}v_jB_k-\frac{1}{\alpha}D^av_av^i\right]+\rho v^i\,.
	\label{eq:MHDcurrent}
	\end{split}
	\end{align}
	In practice, we set $\sigma = 1$ as (approximately) the highest possible value without reaching stiffness in the evolution equations. In cgs units this corresponds to $\sigma\sim 1.6\times 10^4\,\text{s}^{-1}$. 
	
\begin{figure}
	\centering
	\includegraphics[width=0.48\textwidth]{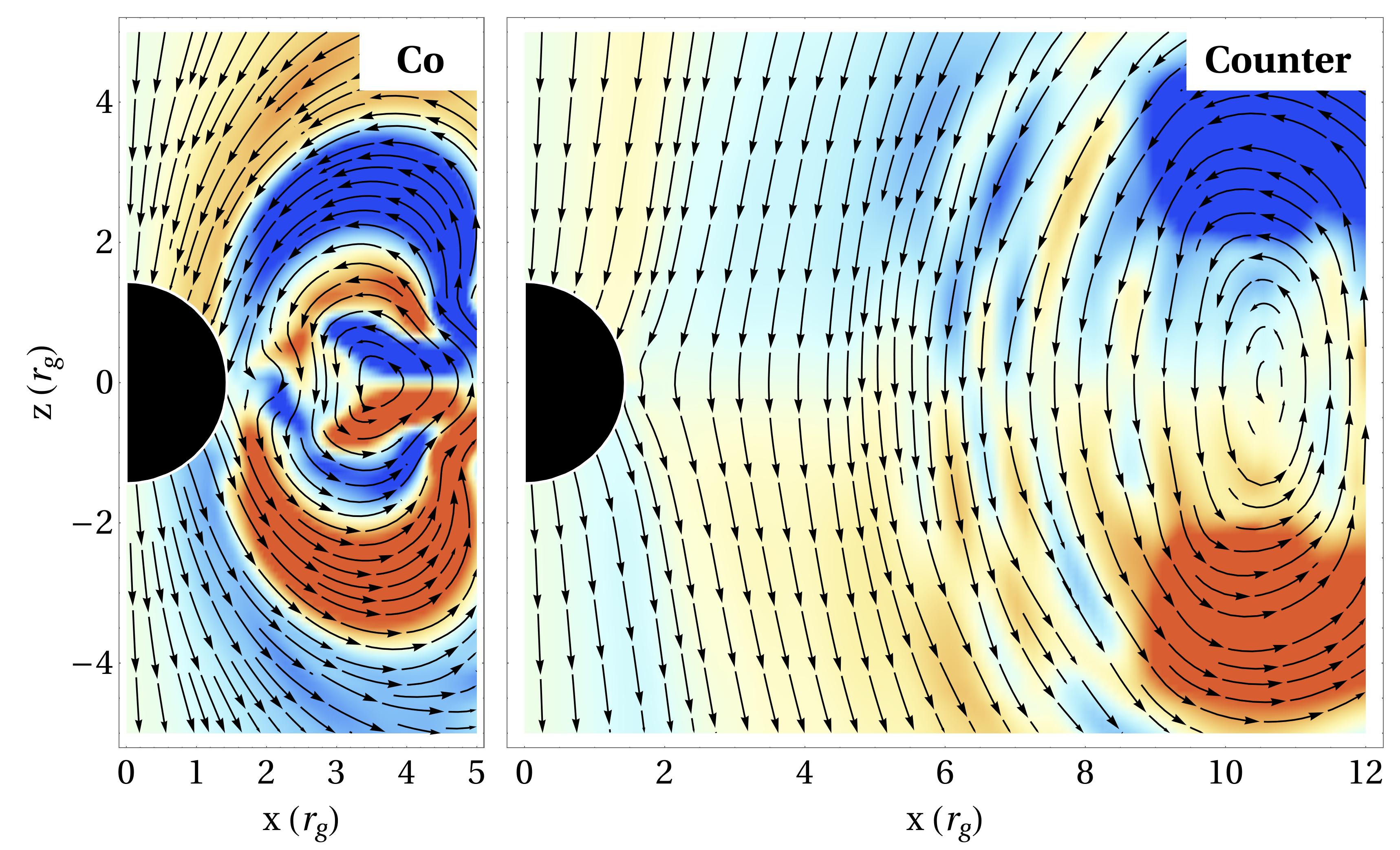} 
	\vspace{-16pt}
	\caption{Zoom of the innermost region of the computational domain showing poloidal streamlines. The colors show the values of the toroidal field $B^T$ (\textit{blue} negative, \textit{red} positive) for an (exemplary) initialization of the co-rotating AD (\textit{left}) and the counter-rotating AD (\textit{right}) for $v^r=0$.}
	\label{fig:Init_Configs}
\end{figure}

	In the UAL (beyond the ISCO) we impose a uniform radial accretion speed, $v_0$ and nearly Keplerian angular velocity,
	\begin{align}
	\begin{split}
	v^r=v_0\,,\qquad\text{and}\qquad v^\phi(r_c)=\pm\frac{\sqrt{M}}{r_c^{3/2}\pm a \sqrt{M}}\,.
	\label{eq:VelocityOuter}
	\end{split}
	\end{align}
	The $+$ and $-$ signs represent co-rotating and counter-rotating disc systems, respectively. Physically, this simple prescription for the accretion speed can be motivated by the fact that in a standard \cite{Shakura_Sunyaev_1973A&A....24..337} disc, the accretion speed is related to the $\alpha_{\textsc{ss}}$ parameter, the disc half-thickness, $h$, and the Keplerian speed, $v_{\rm K}\simeq (r_g/r)^{1/2}c$, through $v_0\simeq \alpha_{\textsc{ss}}(h/r)^2 v_{\rm K}$. In our case, we approximate $\chi_{\rm h}:=h/r\lesssim 0.5$ (a thin disc would require $\chi_{\rm h} \ll 1$, but simulating very thin discs is numerically challenging), and evaluate the Keplerian speed at $r=r_{\rm ISCO}$. Employing the typical value $\alpha_{\textsc{ss}}\simeq 0.1$, we obtain for counter-rotating discs 
\begin{align}	
	v_0\lesssim 0.02(\alpha_{\textsc{ss}}/0.1)(\chi_{\rm h}/0.5)^2 .\label{eq:v_0max}
\end{align}
	We prescribe a toroidal current $J^\phi_{\rm disc}$ in order to create an electric current in the LAL that supports the \textit{ad hoc} magnetic structure (i.e. a series of concentric magnetic flux tubes with alternate polarity). This current is driven during the initialisation and also during the whole evolution of the system. Otherwise, the loop structure in the AD and elsewhere is distorted, rapidly dissipated and eventually destroyed. It is located underneath the UAL, precisely in the region $r_c>r_{\rm ISCO}$ and $\left|z\right|\leq h_\textsc{lal}$:
	\begin{align}
	\begin{split}
	J^\phi_{\rm disc}\left(r_c,t\right)=J_0\times\cos\left(\pi\frac{r_c-r_{\rm ISCO}+t v_0}{l}\right)\times\frac{\alpha}{\sqrt{g}\sqrt{g_{rr}g_{\phi\phi}}}
	\label{eq:InitCurrent}
	\end{split}
	\end{align}
	Here, $l$ denotes the length of the loop along $r_c$. In our setup, each model is characterised by three parameters defining the magnetic structure and accretion of the loops in the AD: The loop length $l$, the loop height within the disk, $h$, which reaches the maximal vertical extension of the UAL, and the (uniform) accretion speed $v_0$. The latter is applied to shift $J^\phi_{\rm disc}$ inwards in time and to prescribe a plasma velocity for the current in the UAL (\ref{eq:MHDcurrent}). We employ $J_0=0.1$ for all the presented setups. By construction, the total magnetic flux accumulated by one tube can differ for different loop dimensions. Also, we use smooth transitions functions blending the different regions composing our BH/AD system into one-another at their interfaces. Outside of the disc and plunging region, the time evolution is fully force-free (uncolored regions in Figure~\ref{fig:discsetup}). Figure~\ref{fig:Init_Configs} shows the (exemplary) initialisation of a loop system by the prescribed current (\ref{eq:InitCurrent}) for the simplified case without accretion ($v^r=0$) and a force-free plunging region.

%--------------------------------------------------------------------
%--------------------------------------------------------------------
\subsection{Toy-model accretion disc setup}
\label{sec:disc_setup}
%--------------------------------------------------------------------
%--------------------------------------------------------------------

The transport of magnetic flux in the plunging region (cf. Figure~\ref{fig:discsetup}, section~\ref{sec:loop_systems}) is a central ingredient in our models. Without imposing a \textit{fluid velocity} corresponding to a geodesic in-fall motion in the \textit{ad-hoc} resistive MHD current (\ref{eq:MHDcurrent}), the magnetic loops of zero net flux are rapidly destroyed. The governing equations of force-free electrodynamics can be written in terms of the evolution of the Poynting flux $S^j$ and the field energy density $e$, rather than the electromagnetic fields (the latter as introduced in section~\ref{sec:force-free}). The time evolution of $e$ reads
\begin{align}
\left(\sqrt{\gamma}e\right)_{,t}&=\:\left(\sqrt{\gamma} S^j\right)_{,j}-\alpha\sqrt{\gamma} D_i J^i+\mathcal{S}_t\label{eq:energy_evolution}.
\end{align}
Here, $\mathcal{S}_t$ corresponds to source terms induced by the geometry or the cleaning of numerical errors. The term $\alpha\sqrt{\gamma}D_i J^i$ vanishes if there is no dissipation, otherwise, it accounts for the losses due to ohmic heating. 
The electromagnetic fields surrounding a resistance wire \citep[cf. Figure~27-5,][]{Feynman2011} are comparable to those imprinted by the innermost current in the plunging region of our model. In the UAL and LAL, the electric field $\mathbf{D}$ is (in part) aligned with the current $J^\phi_{\rm disc}$, such that 
\begin{align}
D_i J^i=D_\phi J^\phi_{\rm disc}\ne 0.
\end{align}
In a stationary magnetic loop structure, the left-hand side of (eq.~\ref{eq:energy_evolution}) vanishes, and the Poynting flux $S^j$ points towards the center of the flux tubes. This is counter-balanced by the ohmic heating term. If $J^\phi_{\rm disc}$ is replaced by the force-free current $\alpha I^\mu_{\textsc{ff}}$ (\ref{eq:FFCurrent}), such heating is prevented by conditions~(\ref{eq:FFCondI}) and~(\ref{eq:FFCondII}):
\begin{align}
D_i J^i=\alpha D_i I^i_{\textsc{ff}}= 0.
\end{align}
The Poynting flux is no longer balanced and the loop becomes a sink for the energy of the force-free field. As the rate of dissipation is given by $\nabla\cdot\mathbf{S}$, a quenching of the loop (i.e. steeper gradients) by the subsequently accreting structure will enhance energy dissipation.

More generally, \citet{Gralla2014} revisit and extend arguments by \cite{MacDonald1982} proving that \textit{a contractible force-free region of closed poloidal field lines cannot exist in a stationary, axisymmetric, force-free Kerr black hole magnetosphere}. Also, \textit{a stationary, axisymmetric, force-free, magnetically dominated field configuration cannot possess a closed loop of poloidal field lines} \citep{Gralla2014}. Though the presented simulations are fully dynamic across all spatial dimensions, closed magnetic loops - wrapped to toroidal flux tubes - or 'closed zones' of field lines connecting the BH to itself seem to be merely transient phenomena in the force-free domain. We conclude that a current for the transport of magnetic field lines through the plunging region, especially the addition of a suitable non-force-free domain along the equator, is essential to sustain closed magnetic loops in their advection towards the BH.

%--------------------------------------------------------------------
%--------------------------------------------------------------------
\section{Simulations}
\label{sec:simulations}
%--------------------------------------------------------------------
%--------------------------------------------------------------------

For all the simulations we employ our own implementation of a GRFFE code (see appendix~\ref{sec:augmented_system} for an overview of the employed conservative scheme) in the framework of the \texttt{Einstein Toolkit}\footnote{\url{http://www.einsteintoolkit.org}} \citep{Loeffler2012}. The \texttt{Einstein Toolkit} is an open-source software package utilizing the modularity of the \texttt{Cactus}\footnote{\url{http://www.cactuscode.org}} code \citep{Goodale2002a}, which enables the user to specify so-called \texttt{thorns} in order to set up tailored simulations. The spacetime is integrated in time using the \texttt{ML\_BSSN}\footnote{\url{http://www.cct.lsu.edu/~eschnett/McLachlan/}} implementation of the BSSN formalism \citep{Brown2009}. We make use of a variety of open-source software, such as the event horizon finder \texttt{AHFinderDirect} \citep{Thornburg2004}, the extraction of quasilocal quantities \texttt{QuasiLocalMeasures} \citep{Dreyer2003}, and the efficient \texttt{SummationByParts} \citep{Diener2007}.

We have performed numerical simulations of the accretion of magnetic loops onto rapidly spinning black BHs using the rescaled \cite{Liu2009} spacetime initial data \citep[as employed also by][]{Mewes2016} for a mass $M=1$ and a reference rotation rapidity $a^*=0.9$. In this paper, we evolve the space-time metric quantities decoupled (i.e. without feedback) from the electromagnetic fields. This decoupling facilitates the comparison to the previous results with a similar (though axisymmetric) setup \citep{Parfrey2015}. The metric quantities only experience a \textit{numerical relaxation} from the initially set up values to the chosen mesh and gauge during the first $\Delta t=50\,r_g$. Throughout the entire simulation ($\Delta t=1024r_g$), $a^*$ decreases by $~\sim 1.4\%$ of its initial value for numerical reasons. Values of $a^*$ very close to 1 are numerically challenging for our code, which, by design, evolves the space-time metric quantities. \citet{Liu2009} are able to obtain a numerically stable evolution, where the metric quantities do not evolve by more than $\sim 1\%$ with respect to their initial values, for a BH $a^*=0.99$ during a relatively short period of time employing larger numerical resolution than used here. However, we have performed a comprehensive parameter space coverage running our models for significantly longer times ($\Delta t\simeq 1000\,M$), making it impractical to employ larger numerical resolution than we have used. Hence, we consider cases with $a^*=0.9$. This value is not as close to 1 as would be desirable to address nearly maximally spinning BHs in AGNs, and it is slightly smaller than the value employed by \citet{Parfrey2015}. However, the outward-going Poynting flux is comparable to the nominal radiative efficiency for $a^*\backsimeq 0.9$ \citep{Hawley2006}, and the value we employ suffices to demonstrate the efficiency of the (intermittent) BZ mechanism when there is no net magnetic flux supplied to the BH. Initially, the electromagnetic field is set to zero everywhere. For an initialization period $\Delta t_{\rm init}=250r_g$, we let the numerical code to build up the electromagnetic fields according to the set of currents described in section~\ref{sec:loop_systems}. At this point of the evolution, the spacetime has fully relaxed to its numerical state of equilibrium, and we start our analysis of physical quantities for an interval $\Delta t_{\rm tot}=\left[0r_g,774r_g\right]$.

%--------------------------------------------------------------------
%--------------------------------------------------------------------
\subsection{Numerical setup}
\label{sec:numerical_setup}
%--------------------------------------------------------------------
%--------------------------------------------------------------------

%
\begin{table}
	\centering
	\caption{Overview of the parameters of our models clustered by series (first column). From the second to the last column, we list the orientation of the disc rotation (\textit{Counter}- or \textit{Co}-rotation; annotated with a prefix \textit{C} or \textit{R}, respectively), the loop length, $l$ and the height $h$ of the disk, the height $h_i$ of the supporting current, and the accretion speed $v_0$ outside of the ISCO.}
\label{tab:model_initials}
	{\renewcommand{\arraystretch}{1.5}
		\begin{tabular}{ccccccc}
			\hline
			Series & Model & Orientation & $h$ & $h_i$ & $l$ & $v_0$ \\*[0pt]\hline
			A & C-H2-L1 & Counter & 2.0 & 0.75 & 1.0 & 0.01 \\
			 & C-H2-L2 & Counter & 2.0 & 0.75 & 2.0 & 0.01 \\
			 & C-H2-L25 & Counter & 2.0 & 0.75 & 2.5 & 0.01 \\
			 & C-H2-L3 & Counter & 2.0 & 0.75 & 3.0 & 0.01 \\
			 & C-H2-L4 & Counter & 2.0 & 0.75 & 4.0 & 0.01 \\
			 & C-H4-L1 & Counter & 4.0 & 2.0 & 1.0 & 0.01 \\
			 & C-H4-L2 & Counter & 4.0 & 2.0 & 2.0 & 0.01 \\
			 & C-H4-L25 & Counter & 4.0 & 2.0 & 2.5 & 0.01 \\
			 & C-H4-L3 & Counter & 4.0 & 2.0 & 3.0 & 0.01 \\
			 & C-H4-L4 & Counter & 4.0 & 2.0 & 4.0 & 0.01 \\\hline
			B & C-H4-L2-005 & Counter & 4.0 & 2.0 & 2.0 & 0.005 \\
			 & C-H4-L2-02 & Counter & 4.0 & 2.0 & 2.0 & 0.02 \\
			 & C-H4-L2-04 & Counter & 4.0 & 2.0 & 2.0 & 0.04 \\\hline
			C & R-H1-L2 & Co & 1.0 & 0.5 & 2.0 & 0.01 \\
			 & R-H1-L3 & Co & 1.0 & 0.5 & 3.0 & 0.01 \\
			 & R-H1-L4 & Co & 1.0 & 0.5 & 4.0 & 0.01 \\\hline
	\end{tabular}}
\end{table}

All shown simulations are conducted in a 3D box of dimensions $\left[2056 r_g\times2056 r_g\times2056 r_g\right]$ with a grid spacing of $\Delta_{x,y,z}=64r_g$ on the coarsest grid level. We employ eleven additional levels of mesh refinement, each increasing the resolution by a factor of two and encompassing the central object. In order to increase resolution in the funnel and disc regions, several levels are stretched in the equatorial direction. Each model (see Table~\ref{tab:model_initials}) is evolved for a period of $t=1024r_g$, or approximately $\sim 50$ revolutions of the central object (corresponding to $a^*=0.9$). On the finest refinement level, we employ a CFL of 0.25, while on coarser levels the timestep needs to be limited to $\delta t\leq1.0$ due to instabilities introduced by the BSSN
gamma driver \citep{Schnetter2010}.

Since all characteristics at the BH horizon point inwards \citep{Faber2007}, information does not propagate from the interior of the horizon outwards. Thus, for numerical convenience, we may reset all variables inside the outer horizon for numerical convenience. Otherwise, close to the BH singularity, the FF equations develop large numerical errors, which may result in the failure of the method. A similar strategy has been employed, e.g. in \citet{Mewes2016}.

In order to ensure the conservation properties of the algorithm, it is critical to employ {\it refluxing} techniques, correcting numerical fluxes across different levels of mesh refinement \citep[see, e.g.][]{Collins2010}. Specifically, we make use of the thorn \texttt{Refluxing}\footnote{Refluxing at mesh refinement interfaces by Erik Schnetter: \url{https://svn.cct.lsu.edu/repos/numrel/LSUThorns/Refluxing/trunk}} in combination with a cell-centered refinement structure \citep[cf.][]{shibata2015}. We highlight the fact that employing the refluxing algorithm makes the numerical code $2-4$ times slower for the benefit of enforcing the conservation properties of the numerical method (specially of the charge). Refluxing also reduces the numerical instabilities, which tend to develop at mesh refinement boundaries.

%--------------------------------------------------------------------
%--------------------------------------------------------------------
\subsection{Energy outflow}
\label{sec:energy_outflow}
%--------------------------------------------------------------------
%--------------------------------------------------------------------

%
\begin{figure*}
	\centering
	\includegraphics[width=0.99\textwidth]{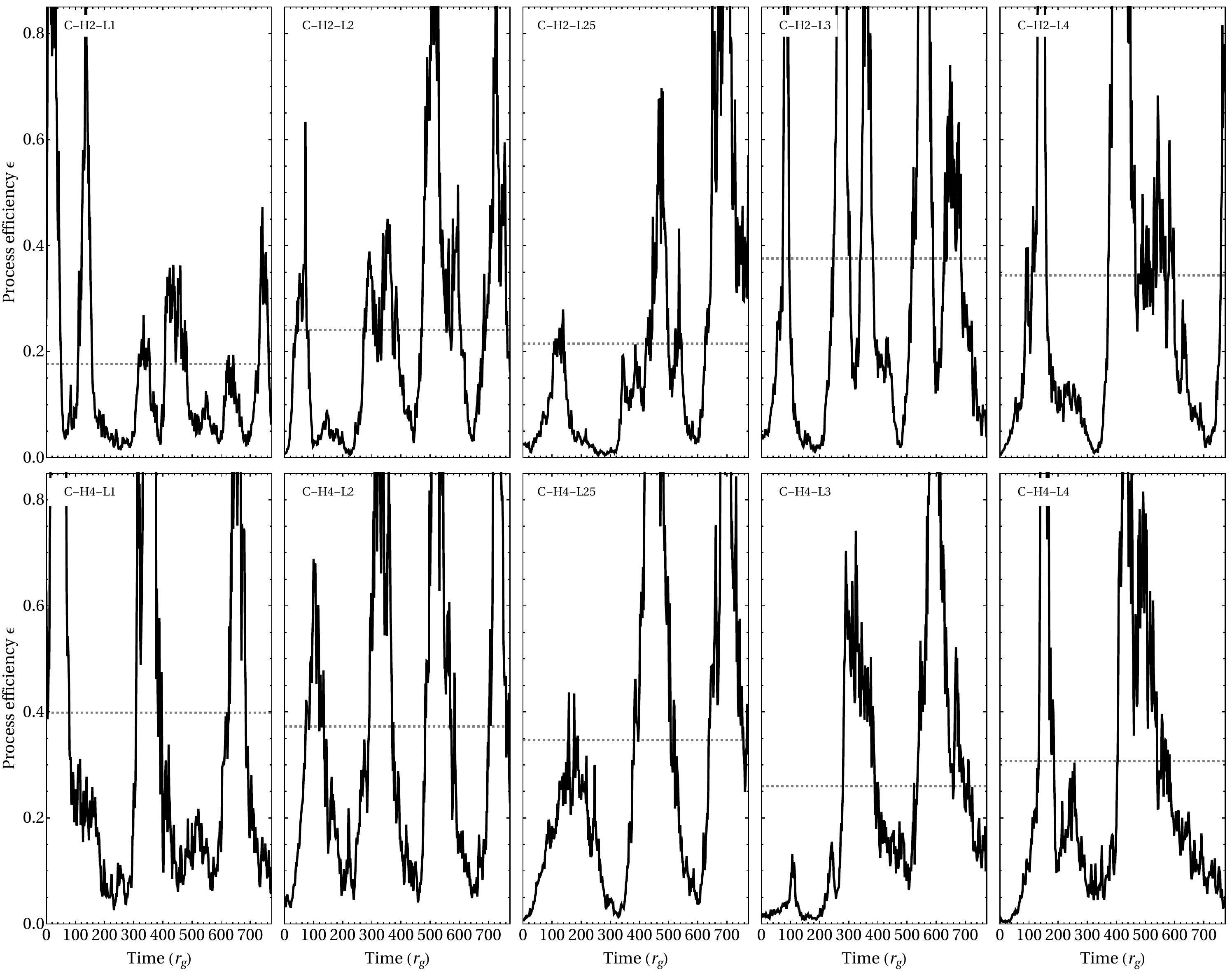} 
	\vspace{-8pt}
	\caption{Efficiency of the BZ process $\epsilon$ (eq.~\ref{eq:process_efficiency}) during the evolution of selected models of counter-rotating discs (cf. Table~\ref{tab:model_initials}). The process efficiency depends on the structure of the magnetic loops in the AD, specifically on the chosen loop scale height and length, in a complex and non-linear way. The average efficiency $\bar\epsilon$ of each model over the entire simulation period $\Delta t_{\rm tot}$ is indicated by a gray dashed line.}
	\label{fig:counter_energy}
\end{figure*}

Efficient energy extraction of the reducible energy from rapidly spinning BHs via the BZ process happens if the field line angular velocity, $\Omega_{\rm F}$, equals half the BH angular frequency, i.e.
$\Omega_{\rm F}=\Omega_{\rm BH}/2$ \citep{Blandford1977}. Under this optimal condition, a second order accurate estimation\footnote{We note that in the preceding work \citet{Mahlmann2018}, the BZ power has a misplaced factor of $M^2$  in eq.~42 of \cite{Mahlmann2018}.} of the luminosity of the BZ process reads \citep{Tchekhovskoy2015}
\begin{align}
\mathcal{L}_{\rm BZ}\approx \frac{1}{24\pi^2}\Omega_{\rm BH}^2\Phi^2=\frac{1}{96\pi^2}\left(\frac{ \Phi a^*}{r_+}\right)^2.
\label{eq:BZ_Luminosity}
\end{align}
The factor $1/96\pi^2\approx 10^{-3}$ corresponds to the split-monopole BH magnetosphere\footnote{In Heaviside-Lorentz geometrised units \citep[cf.][]{Mahlmann2019}, which differ by a factor $1/\sqrt{4\pi}$ when compared to Gaussian cgs units \citep[as displayed, e.g., by][]{EHT2019v}.}, and depends weakly on the field geometry \citep[cf.][]{Tchekhovskoy2010}. $\Phi$ denotes half of the absolute magnetic flux through the BH horizon. Our models do not set any initial magnetic field close to the BH horizon, only attached to the AD. Hence, in order to compute an estimate for the BZ luminosity that can be used as a normalization of our results, we make the following assumption: The entire poloidal magnetic flux of a tube detaching from the AD would ideally thread the BH horizon. Thus, we integrate the vertical flux through the equatorial plane ($r,\phi\in\left[r_{\rm ISCO}+l/2\,,\,r_{\rm ISCO}+l\right]\times\left[0,2\pi\right]$) for the last loop outside of the ISCO, in order to derive an upper limit to $\Phi$:
\begin{align}
\Phi=\iint_{r\phi}\left|B^\theta\right|\,\text{d}A_\theta.
\end{align}
Here, $\text{d}A_i$ denotes a suitable area element. In actuality, part of this flux may be lost due to magnetic reconnection and may never end up touching the BH horizon. Following \citet{Parfrey2015}, we define the process luminosity $\mathcal{L}_{\rm LA}$ during the accretion of magnetic loops as the surface integral of the outgoing Poynting flux $S^r_+$ over the BH horizon, hence, only including the sum of positive contributions to the total energy flow:
\begin{align}
\mathcal{L}_{\rm LA}=\iint_{\theta\phi}S_+^r\,\text{d}A_r
\end{align}
The Poynting flux $S^r_+$ is derived from the corresponding components of the energy-momentum tensor \citep{Komissarov2004},
\begin{align}
S^i=T^i_{\hspace{4pt}t}&=\:-\frac{1}{\alpha}e^{ijk}E_j H_k.
\end{align}
The process efficiency relating the energy extracted by the in-fall of magnetic loops onto the central object compared to an optimal BZ powered energy extraction then reads:
\begin{align}
\epsilon=\frac{\mathcal{L}_{\rm LA}}{\mathcal{L}_{\rm BZ}}.
\label{eq:process_efficiency}
\end{align}
Besides the instantaneous variation of the efficiency shown in Figures~\ref{fig:counter_energy} and~\ref{fig:co_energy}, it is important to asses whether the accretion of loops with zero net magnetic flux drives, on average, a significantly luminous outflow. For that we may compute the time averaged efficiency, $\bar{\epsilon}$, over the whole computed time $\Delta t_{\rm tot}$. However, we realize that in many models one needs to wait for two or three cycles before some quasi-periodic behavior takes place. Since evolving our 3D models much longer is prohibitive, we consider an alternative measurement of the average efficiency. Namely, we quantify the average efficiency $\left\langle\epsilon\right\rangle$ during the accretion period $\Delta t_{\rm acc}=l/v_0$ of the final accretion cycle in the computed time.

%--------------------------------------------------------------------
%--------------------------------------------------------------------
\subsubsection{Counter-rotating accretion disc}
\label{sec:Counter-rotatingAD_efficiency}
%--------------------------------------------------------------------
%--------------------------------------------------------------------

%
\begin{figure}
	\centering
	\includegraphics[width=0.425\textwidth]{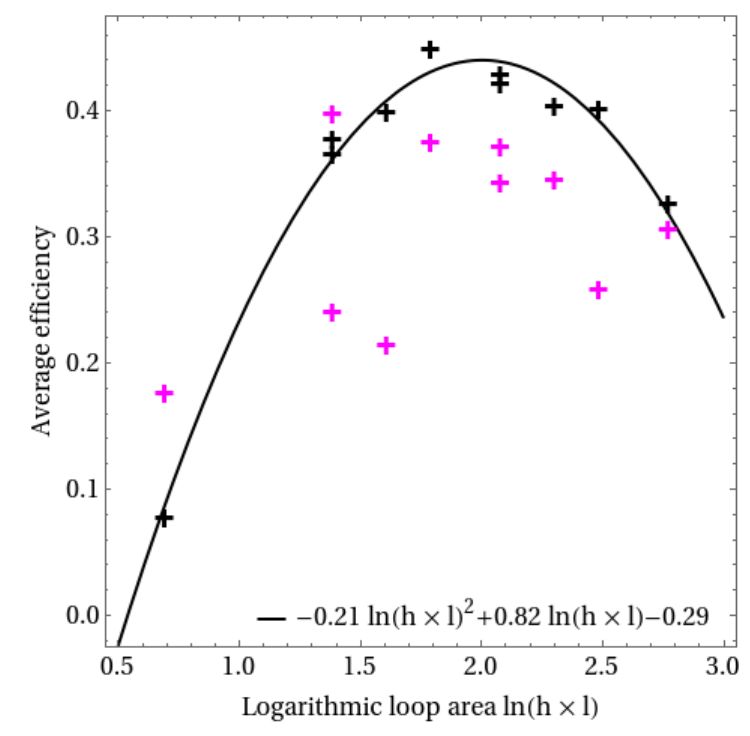} 
	\vspace{-8pt}
	\caption{Average efficiency of the BZ energy extraction in selected models of counter-rotating discs (series A, cf. Figure~\ref{fig:counter_energy} and Table~\ref{tab:model_initials}). The average efficiencies against the logarithmic loop area are depicted by crosses. The average efficiencies $\bar{\epsilon}$ over the simulation period $\Delta t_{\rm tot}$ are shown in magenta, the efficiency of the last accreted loop $\left\langle\epsilon\right\rangle$ for each model in black, a quadratic fit function is depicted by a respective line.}
	\label{fig:Counter_Average}
\end{figure}

The simulation time for all models of series A (see Table~\ref{tab:model_initials}) is sufficient to detach several loops from the AD and model their plunge onto the central BH. Figure~\ref{fig:counter_energy} visualizes the evolution of the process efficiency $\epsilon$. Peaks of efficient outgoing Poynting fluxes can be seen throughout all models of the series. While the calculated peak efficiency of $\epsilon_P\approx 85\%$ is similar for all the shown models, consecutive peaks often differ in shape and fine structure. The efficiency peaks are related to the structure of magnetic loops in the AD in a complex and non-linear way. They do not follow the simplistic expectation according to which, after the accretion of half of a complete magnetic loop, an efficiency peak develops while low efficiency occurs only for times in between of two consecutive loops of alternated polarity (when the magnetic flux threading the BH horizon is closest to zero). Furthermore, the temporal width of the peaks is not a one-to-one map of the time need to accrete half of a complete magnetic flux tube from the accretion disc, namely, $\Delta t_{\rm acc}$. Indeed, the consecutive episodes of efficient energy extraction show, in many cases, a lot of substructure and both the peak shape and $\epsilon_P$ notably differ from peak to peak (e.g. see models C-H2-L2, C-H2-L3, C-H4-L4). This contrasts with the results of \cite{Parfrey2015}, where the high-efficiency pulses of the single model (shown in their Figure~2) are very regular and reach nearly the same value of $\epsilon_P\simeq 0.75$ in all cases. Only during the first peak, some transitory relaxation of the initial conditions is observed in their 2D models. We attribute the differences to the complex 3D dynamics and to the fact that the loop cross-sectional size is a factor of foremost importance shaping the efficiency of energy extraction.

%% Effect of the vertical size:
Models with small length of the loops, $l=1r_g$ (C-H2-L1 and C-H4-L1), convert magnetic flux into a Poynting dominated energy outflow less efficiently. The difference in the vertical extension of the AD between models C-H2-L1 and C-H4-L1 induces significant differences in the regularity of the high-efficiency episodes. Three relatively regular episodes of high efficiency (at the peak, $\epsilon_P> 0.8$) with duration $\Delta t\sim 100\,r_g$ follow each other in model C-H4-L1, while only two peaks with $\epsilon_P> 0.8$ and duration $\Delta t\sim 100\,r_g$ are irregularly distributed in $\Delta t_{\rm tot}$ for model C-H2-L1. In both cases, efficient episodes are followed by less powerful cycles. During the absence of powerful outflows, we observe that the structure of wound up field lines threading the BH horizon fails to open up to high vertical extensions (see discussion in section~\ref{sec:loopefficiency}). The rapid release of flux tubes of shorter length is also imprinted onto the shown efficiency curves by an increased small-scale variability due to more incoherent flux structures arriving at the BH horizon.

For series A, average efficiencies during the accretion of one (or two, in case of the models of loop length $l=1r_g$) magnetic loops are shown in Figure~\ref{fig:Counter_Average} (black symbols) as a function of the logarithm of $l\times h$, which is proportional to the cross-sectional area (in the poloidal plane) of the magnetic flux tubes setup in the AD. In this representation, one can identify a range of optimal loop cross-section areas for which the average efficiency is nearly maximal, $\left\langle\epsilon\right\rangle\approx 0.36-0.43$. This range is rather broad and corresponds to models with very similar loop cross-sectional area, namely, C-H2-L2 and C-H4-L1 as well as C-H2-L4 and C-H4-L2. For very small and very large loop areas, $\langle \epsilon \rangle$ drops to lower values. We stress that $\langle \epsilon \rangle$ cannot be interpreted using independently $l$ or $h$ as parameters. Only the combination of both (in the form $h\times l$) permits finding some empirical correlation between the geometrical properties of the loops and the process efficiency. After testing many different possibilities, we find that the average process efficiency can be fit by (see black line in Figure~\ref{fig:Counter_Average})
\begin{equation}
\left\langle\epsilon\right\rangle \simeq -0.21\left[\ln(h\times l)\right]^2+0.82\,\text{ln}{(h\times l)}-0.29\, .
\label{eq:efficiency_size_conter}
\end{equation}

We also display the average efficiency over the whole computed time, $\bar\epsilon$, in Figure~\ref{fig:Counter_Average} (magenta symbols). The dependence on the surface area of the loops found for $\langle \epsilon\rangle$ is much less evident for $\bar\epsilon$. This is due to the fact that during the accretion of the first loop the dynamics in the BH magnetosphere is still rather violent and an approximately steady state has not been formed. We note that a qualitatively similar difference between the first loop of the series and the subsequent ones was also found by \cite{Parfrey2015}. This behavior justifies our choice of measuring the average efficiency over the last loop accreted during $\Delta t_{\rm tot}$, $\langle\epsilon\rangle$. It provides a cleaner interpretation of the dependence of results on the model parameters.

%--------------------------------------------------------------------
%--------------------------------------------------------------------
\subsubsection{Co-rotating accretion disc}
%--------------------------------------------------------------------
%--------------------------------------------------------------------

%
\begin{figure}
	\centering
	\includegraphics[width=0.45\textwidth]{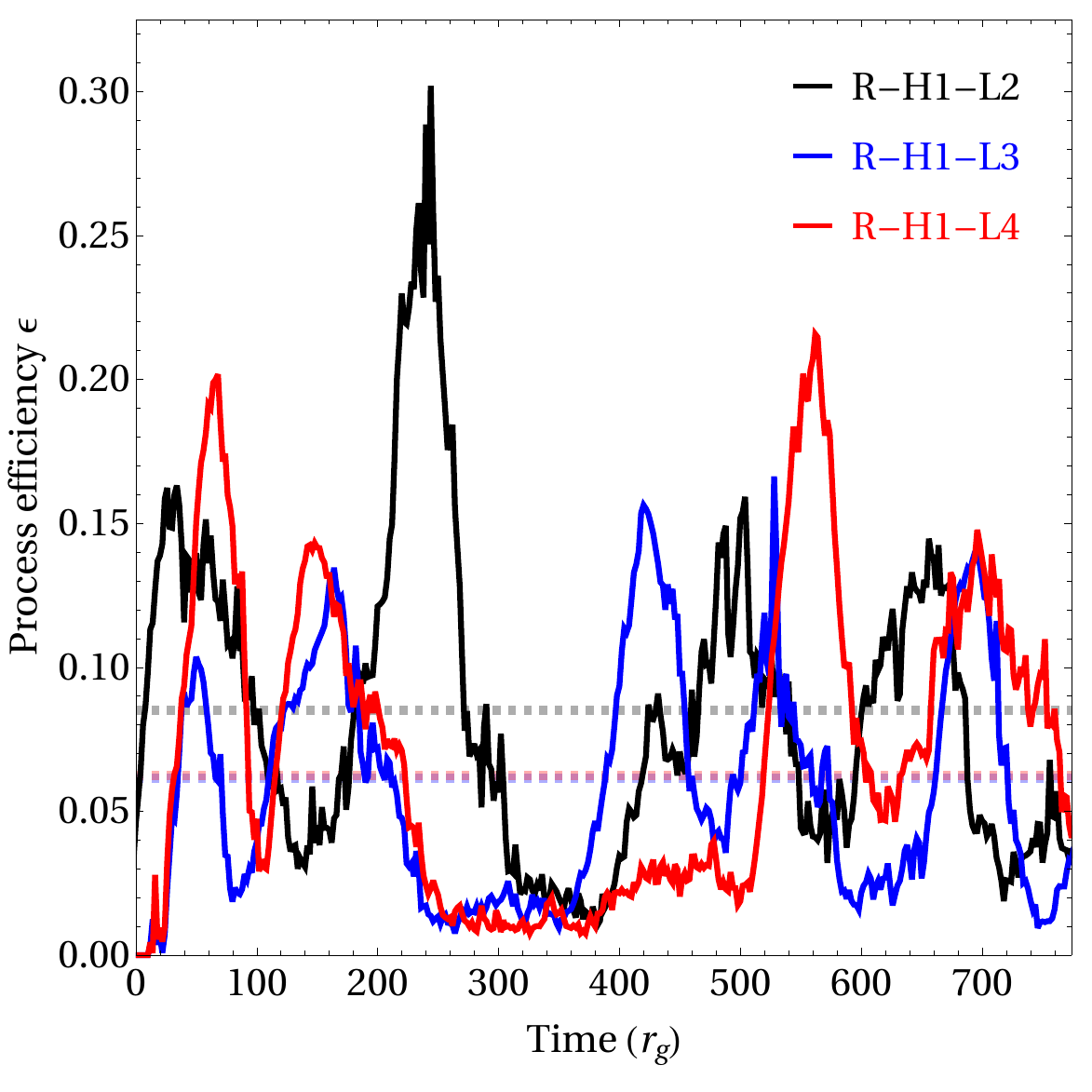}
	\vspace{-12pt} 
	\caption{Same as Figure~\ref{fig:counter_energy} but for the co-rotating models of series C in Table~\ref{tab:model_initials}. Both, the instantaneous and average efficiencies are notably smaller than in the case of counter-rotating discs.}
	\label{fig:co_energy}
\end{figure}

Figure~\ref{fig:co_energy} visualizes the evolution of the instantaneous efficiency $\epsilon$ of the models in series C (see Table~\ref{tab:model_initials}). All of these models show significantly lower process efficiencies (with $\bar{\epsilon}\approx 0.06-0.09$) than the setups of series A. Like in many counter-rotating models, the pulse pattern shows complex substructures and different peak heights. Though the loop length $l$ can be recognized in the overall periodicity, the average efficiency $\langle\epsilon\rangle$ decreases for the models with the larger loop length R-H1-L3 and R-H1-L4. This happens because of the existence of longer \textit{quiescent} periods between some of the peaks, where the instantaneous efficiency nearly drops to zero (e.g. between $\simeq 250r_g$ and $\simeq 500r_g$ for model R-H1-L4). Figure~\ref{fig:co_energy} also shows that each accretion cycle for these models has two peaks of $\epsilon_P\approx 0.2$ with a drop in efficiency in between them.

Due to the large BH spin, the location of $r_{\rm ISCO}$ is very close to the BH horizon in case of co-rotating ADs. We are, hence, faced with two important challenges: i) The disc height $h$ has to be chosen such that the disc does not become excessively thick in the vicinity of the BH. Comparatively, the disc height-to-cylindrical radius ratio, $h/r_{\rm ISCO}$ is appreciably smaller for counter-rotating discs ($r_{\rm ISCO}\simeq 8.7r_g$) than for co-rotating discs ($r_{\rm ISCO}\simeq 2.3r_g$) with the same height. In practice, this fact introduces a strong distortion of the loop shape in co-rotating models. This distortion blurs the measurement of the process efficiency. ii) Our \textit{ad hoc} setup induces an additional far-field energy flow into our domain (see appendix~\ref{sec:faraday_disc}). The models of series C (Figure~\ref{fig:co_energy}) are chosen such that these caveats do not affect the energy flows at the BH horizon.

%--------------------------------------------------------------------
%--------------------------------------------------------------------
\subsection{Field structure}
\label{sec:field_structure}
%--------------------------------------------------------------------
%--------------------------------------------------------------------

%--------------------------------------------------------------------
%--------------------------------------------------------------------
\subsubsection{Counter-rotating accretion disc}
%--------------------------------------------------------------------
%--------------------------------------------------------------------

%
\begin{figure*}
	\centering
	\includegraphics[width=0.99\textwidth]{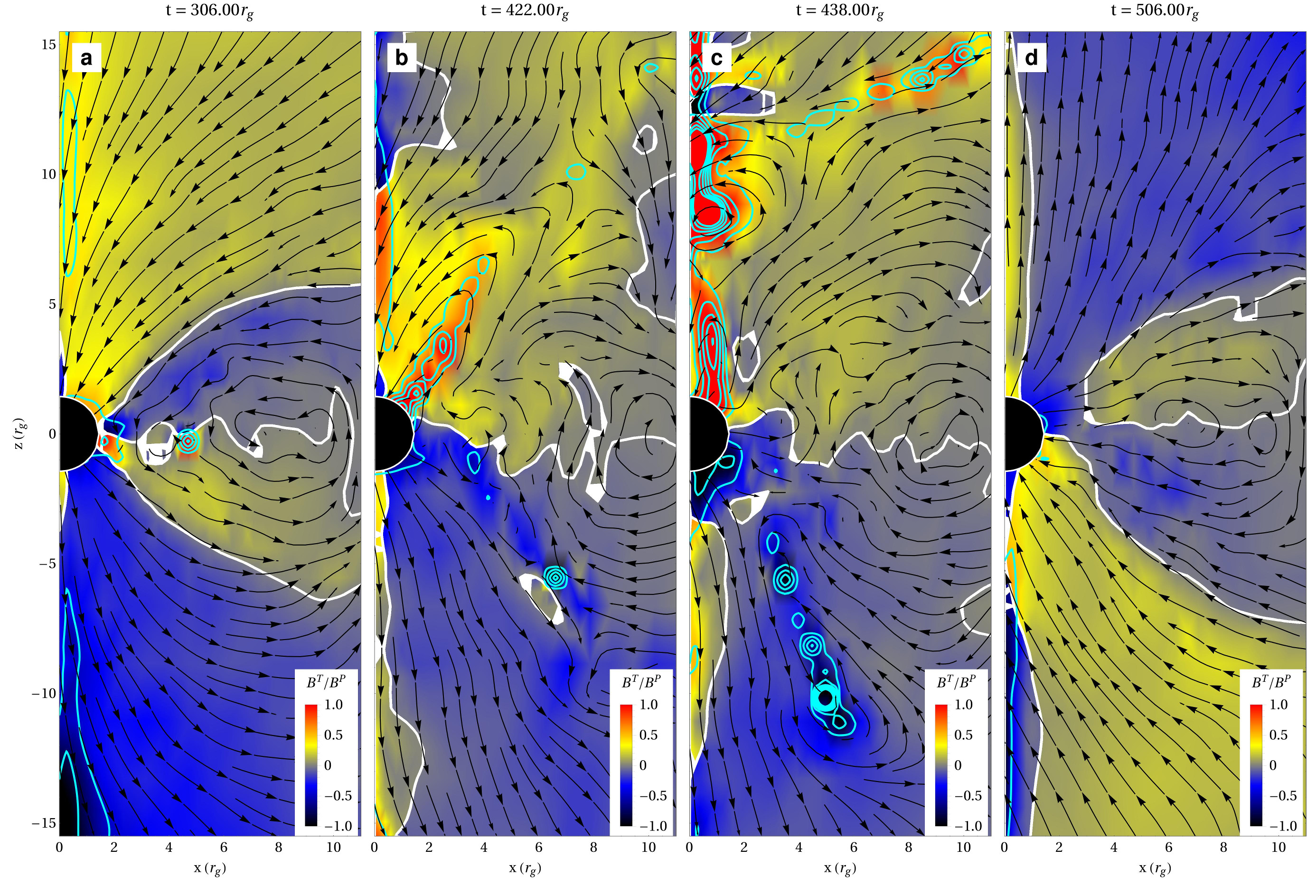} 
	\vspace{-12pt}
	\caption{Evolution of the toroidal magnetic field dominance between two subsequent peaks of efficient energy extraction (model C-H4-L2). The colour scaling represents the strength of the toroidal magnetic field (including its sign) scaled by the magnitude of the poloidal magnetic field. The white lines indicate a change in polarity of the toroidal magnetic field. Cyan contours indicate very strong toroidal field dominance, e.g. as found in plasmoids. The shown data includes refinement levels of resolutions $\Delta_{x,y,z}\leq 0.5r_g$. Episodes of efficient energy extraction (\textbf{\textsf{a}} and \textbf{\textsf{d}}) show recurring, ordered field structures. During the detachment of a magnetic loop from the AD and the change of overall polarization (\textbf{\textsf{b}} and \textbf{\textsf{c}} plots), simulations show small-scale, disordered structures (resembling turbulence) and plasmoids with strong toroidal field dominance.}
	\label{fig:counter_bpbr}
\end{figure*}
\begin{figure}
	\centering
	\includegraphics[width=0.475\textwidth]{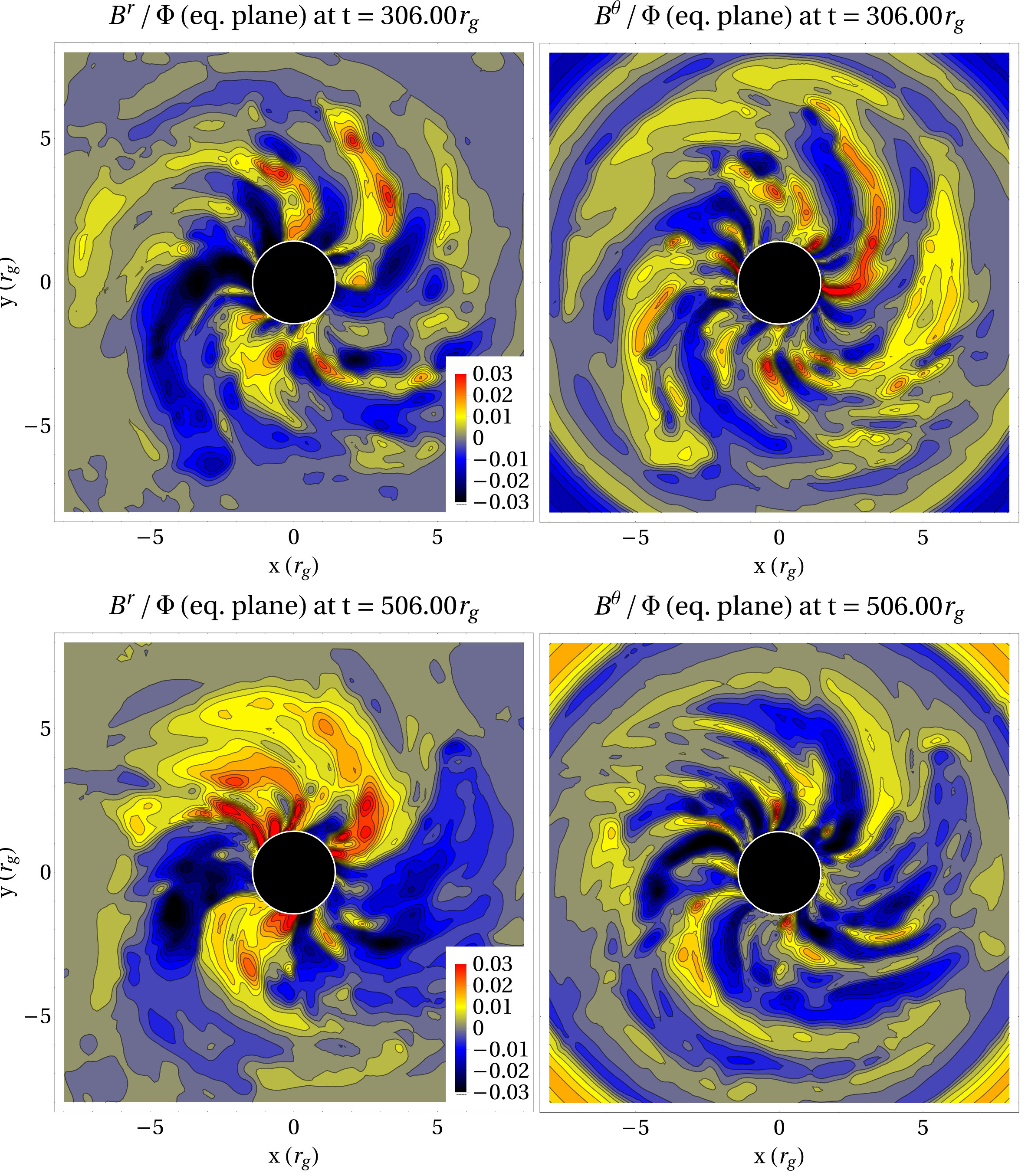}
	\vspace{-15pt}
	\caption{The $B^r$ and $B^\theta$ components in the equatorial plane (model C-H4-L2) show periodic in-spiraling of magnetic flux onto the central BH during two subsequent episodes of efficient energy extraction. The shown data has a grid resolution of $\Delta_{x,y,z}\leq 0.25r_g$. The emerging spiral patterns show 3D effects in the plunging region, which break the axial symmetry.}
	\label{fig:counter_equatorial}
\end{figure}

Once a magnetic loop reaches the inner disc boundary, part of it will start to free-fall onto the BH. This results in the development of structures resembling a \textit{hairpin} \citep[using the naming convention of][]{Beckwith2009} in the plunging region, effectively connecting the BH horizon with the AD by twisted magnetic field lines (see panels \textbf{\textsf{a}} and \textbf{\textsf{d}} of Figure~\ref{fig:counter_bpbr}). The emerging field structure shows a well ordered dipole component, the growth and decrease of which is linked to the energy pulses as depicted in Figure~\ref{fig:counter_energy}. At the same time, the action of strong differential shear in the plunging region (between the ISCO and the BH horizon) opens up magnetic field lines of the previously accreted loop, forming an ordered magnetic field of a parabola-like shape in the jet launching regions above the poles of the central BH (panel \textbf{\textsf{d}}). Though the accretion system supplies tubes of zero net magnetic flux, this structure of ordered magnetic fields is maintained over significant lengths compared both to the loop size and the plunging timescale induced by $r_{\rm ISCO}$.

Once a magnetic flux tube fully disconnects from the AD, several events occur in order to rearrange the magnetic field configuration with the accretion of a new magnetic tube of opposite polarity. During these processes, there is no efficient Poynting induced energy extraction across the BH horizon:
\begin{itemize}
	\item [i)] Establishment of quadrupole and higher multipoles \citep[i.e. emergence of closed loops on either side of the equator, cf.][]{Beckwith2008} small-scale structures resembling turbulence in the boundary between regions of different magnetic polarity (see Figure~\ref{fig:counter_bpbr}\textbf{\textsf{b}} and \textbf{\textsf{c}}). This process comes along with the relaxation of the parabola-like shape in the \textit{jet launching} region, i.e. a biconic region with an approximate half-opening angle $\sim 30^\circ - 45^\circ$ (see Figure~\ref{fig:counter_bpbr}\textbf{\textsf{a}} and \textbf{\textsf{d}}).
%		 .
	\item [ii)] Expelling of large-scale flux structures from the jet launching region and replacing by the opposite polarity fields of the newly accreting tube opening up from the AD (see Figure~\ref{fig:counter_bpbr}\textbf{\textsf{c}}).
	\item [iii)] Evacuation of plasmoids with strong toroidal field dominance along the interface of opposite polarities into the jet launching region and away from the central object (see Figure~\ref{fig:counter_bpbr}\textbf{\textsf{c}}).
\end{itemize}

\begin{figure}
	\centering
	\includemedia[width=1.0\linewidth,
	height=0.915648\linewidth,
	activate=onclick,
	passcontext,
	transparent,
	addresource=images/figure8a.mp4,
	flashvars={flv=images/figure8a.mp4}
	]{\includegraphics[width=1.0\textwidth]{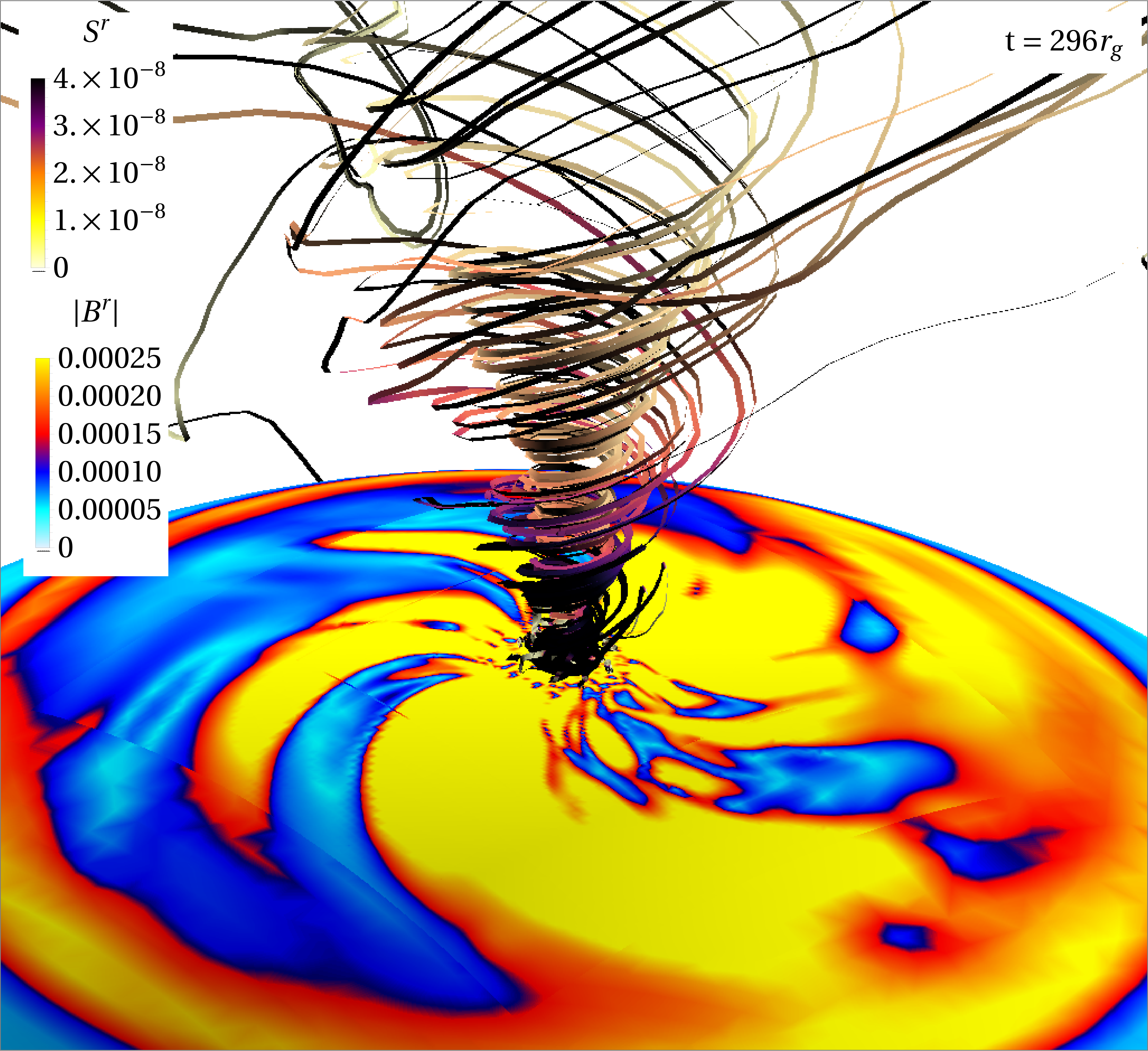}}{player.swf}
	\caption{3D impression of the accretion of one magnetic flux tube onto a rapidly spinning BH ($a^* = 0.9$) in the C-L4-L2 model. The (outgoing) Poynting flux emerging from the BH horizon is visualized by ribbons coloured according to the strength of the associated radial energy flow (CGS units; see the colour scale). The radial magnetic flux (absolute value) is depicted by the density plot, indicating 3D non-axisymmetric effects in the plunging region. During peak outflow, extended helical structures of energy flow build up above the polar regions. Their confinement and strength decreases after peak efficiency. \textit{Click} for animation (only Adobe Reader).}
	\label{fig:3DCounter}
\end{figure}

During the phase of continuous accretion, the magnetic flux through the equatorial plane builds up spiral patterns (Figure~\ref{fig:counter_equatorial}). Such perturbations reflect a loss of both, equatorial and axial symmetry along the equatorial current sheet, while the extended magnetic configuration may still exhibit ordered fields. The loss of symmetry in our perfectly axisymetric intial models is due to both numerical and physical reasons. The hierarchy of nested Cartesian grids mapping an axisymmetric setup may imprint small numerical perturbations on the plunging region, specially, at the boundary between the free-falling plasma and the AD. Episodic reconnection events drive physical perturbations along the equatorial plane also in the plunging region. Remarkably, qualitatively similar reconnection episodes may also break the equatorial symmetry in axisymmetric particle-in-cell simulations \citep{Parfrey_2019PhRvL}. Similar 3D effects have been observed by \citep[][cf. Figure 15]{Beckwith2009} in the context of disconnecting magnetic loops in the accretion funnel of a large-scale magnetic flux system. During phases of efficient energy extraction from the central object (Figure~\ref{fig:counter_energy}), extended helical structures of (outgoing) Poynting flux are formed in the polar regions. Figure~\ref{fig:3DCounter} shows such structures for the C-H4-L2 model at the moment of peak efficiency.

%

%--------------------------------------------------------------------
%--------------------------------------------------------------------
\subsubsection{Co-rotating accretion disc}
%--------------------------------------------------------------------
%--------------------------------------------------------------------

%
\begin{figure*}
	\centering
	\includegraphics[width=0.99\textwidth]{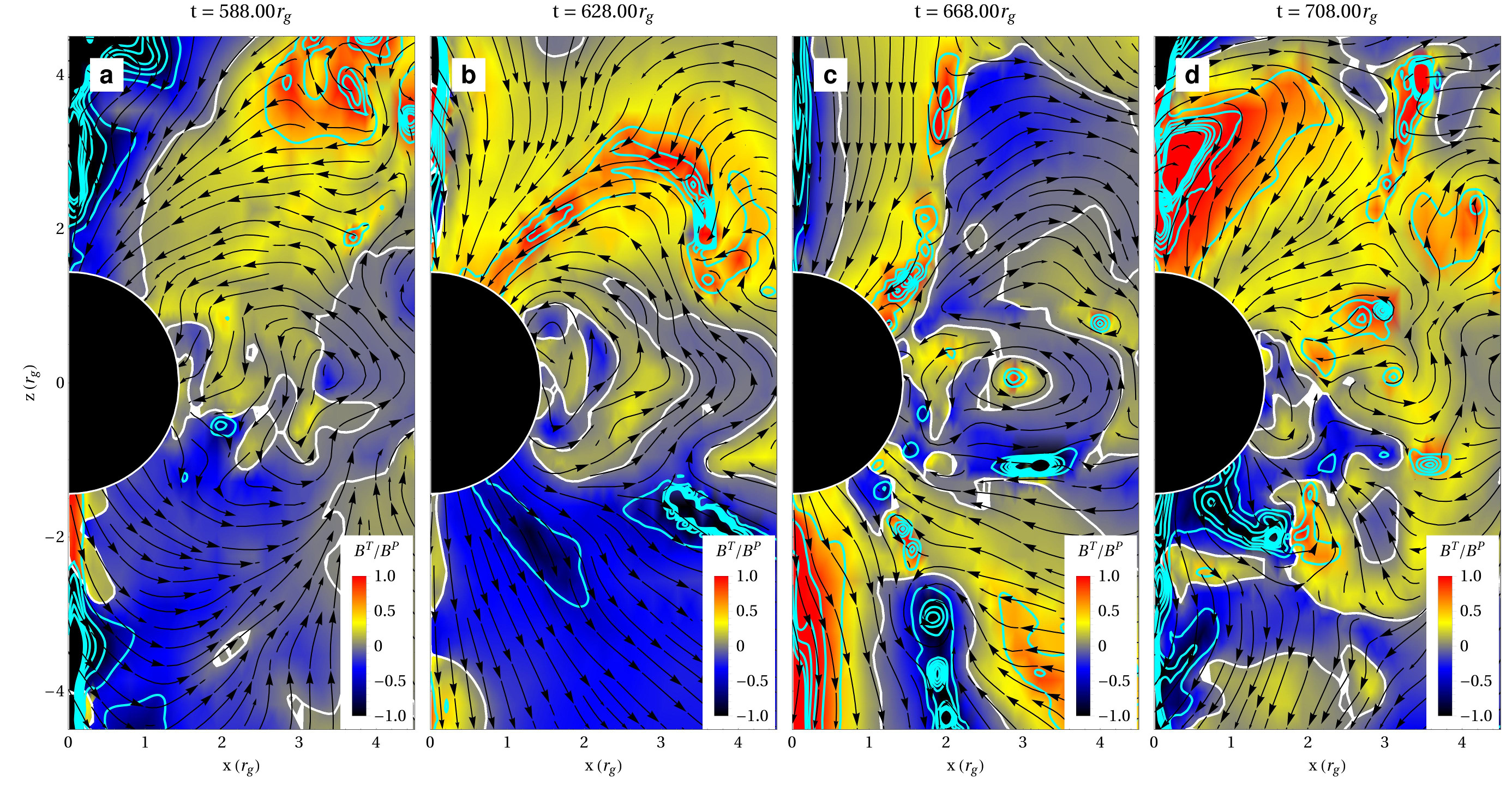} 
	\vspace{-12pt}
	\caption{As Figure~\ref{fig:counter_bpbr}, evolution of the toroidal magnetic field dominance across one peak of efficient energy extraction (model R-H1-L2). The shown data has a refinement resolution of $\Delta_{x,y,z}\leq 0.25r_g$. Episodes of efficient energy extraction (\textbf{\textsf{b}} and \textbf{\textsf{c}}) follow the opening of magnetic loops and the transient development of larger scale flux structures in the jet launching region. Global rearrangements in the disc's funnel trigger the development of plasmoids at the interface of opposing polarities. The connection of the majority of magnetic field lines emerging from the BH to the AD (\textbf{\textsf{a}}) as well as the development of small-scale magnetic fields (\textbf{\textsf{d}}) are paralleled by a vanishing efficiency in the energy extraction.}
	\label{fig:counter_bpbrco}
\end{figure*}

A \textit{stationary}, \textit{axisymmetric} force-free magnetospheres of a rotating BH including both open and closed (co-rotating) field lines anchored in a thin disc was discussed, e.g. by \citet{Uzdensky2005,Mahlmann2018,Yuan2019a,Yuan2019}. In their equilibrium solutions all closed field lines connect the BH horizon to the inner regions of an equatorial (thin) disc up to a cylindrical radius $r_{\rm close}>r_{\rm ISCO}$. The disc also supports open field lines beyond $r_{\rm close}$. The foot-points of both open and closed field lines anchored in the disc rotate with the corresponding angular velocity (see equation~\ref{eq:VelocityOuter}). Closed field lines in this kind of magnetosperic topology allow for the exchange of angular momentum between the BH and the AD, but they do not efficiently extract energy to infinity. \citet{Uzdensky2005} further identifies the possibility of a combination of these closed field lines and open field lines extending to a region far away from the central object in BH/AD systems, effectively extracting part of the energy by the BZ process \citep[see also][on the coexistence of electromagnetic accretion and ejection flows]{Contopoulos2019}. \cite{Parfrey2015} argue that, since magnetic field lines can remain closed only up to $r_{\rm close}$ in axisymmetric magnetospheres, magnetic topologies composed by flux tubes with lengths $l< l_{\rm crit} = r_{\rm close} - r_{\rm ISCO} \sim O(r_g)$ will not produce jets. According to \cite{Uzdensky2005}, the exact location of $r_{\rm close}$ sensitively depends on the problem setup, e.g. the BH spin parameter, and the magnetic flux distribution on the disk. Thus, we also expect that $l_{\rm crit}$ depends on similar factors, in addition, e.g., to the disc's thickness and conductivity, and the radial distribution of the flux tubes in the disc. In the conducted simulations of co-rotating disc models (series C, Table~\ref{tab:model_initials}) we find closed magnetic flux tubes connecting the inner regions of the AD with the BH. These configurations are forming repeatedly, but they are neither axisymmetric nor steady.
Due to the limitations of the idealized setup for prograde AD models (appendix~\ref{sec:faraday_disc}), we cannot reliably separate the contribution of the BZ process from that of the AD in the overall electromagnetic luminosity far away from the BH. It is beyond the scope of this paper to asses the exact value of the critical loop length for various reasons: It would require modifying the AD setup in co-rotating models, so that the disc height be much smaller than $r_{\rm ISCO}$ (as required for thin ADs). Also, our simplified setup for co-rotating ADs is not optimally suited to explore models with larger values of $a^*$ (appendix~\ref{sec:faraday_disc}). Finally, due to the numerical diffusion far away from the BH we would need to increase our resolution significantly in these regions to properly track energy flows towards infinity. These facts have, indeed, limited the numerically explored range of loop widths for prograde discs to values $2\,r_g\le l\le 4\,r_g$ (note that the smallest significant value of the loop width would be $l\approx r_{\rm ISCO}-r_+\approx 0.9\,r_g$ for $a^*=0.9$). Since our model setup differs slightly from that of \cite{Parfrey2015} - e.g. in the smaller value of $a^*$ and in the larger accretion speed - we find that energy extraction is still efficient for $l=2\,r_g$ (see also Section~\ref{sec:loopefficiency}).
Our results are compatible with the existence of a critical loop length, which manifests in our models as a reduction of the efficiency of the energy extraction for our prograde AD models compared to their retrograde counterparts.
Another reason explaining the smaller efficiency of the BZ process in our prograde discs is of numerical origin. In general, maintaining the structural integrity of the AD model proves to be much harder for the co-rotating disc models. Especially in the UAL, the loop structure smears out in the course of the simulations. However, the time evolution shows the following sequence of reoccurring structures:
\begin{itemize}
	\item [i)] Connection of the majority of field lines emerging from the BH to the innermost region of the AD with a vanishing overall energy extraction (see Figure~\ref{fig:counter_bpbrco}\textbf{\textsf{a}}).
	\item [ii)] Opening up of the accreted loop and gradual extension of field lines linking the polar regions to larger scale heights (Figure~\ref{fig:counter_bpbrco}\textbf{\textsf{b}}).
	\item [iii)] Complete opening of the accreted loop and initialization of the rearrangement of the jet launching region. In this phase the peak energy extraction efficiency is attained. This comes along with the formation of larger-scale flux structures above the polar regions, development of plasmoids with strong toroidal field dominance at the interface of different polarizations (Figure~\ref{fig:counter_bpbrco}\textbf{\textsf{c}}).
	\item [iii)] Rearrangement of the fields in the plunging region ensuing the development of extended regions of strong toroidal dominance along the axis of rotation. Decrease in process efficiency (Figure~\ref{fig:counter_bpbrco}\textbf{\textsf{d}}).
\end{itemize}
%

%--------------------------------------------------------------------
%--------------------------------------------------------------------
\section{Discussion}
\label{sec:discussion}
%--------------------------------------------------------------------
%--------------------------------------------------------------------

%--------------------------------------------------------------------
%--------------------------------------------------------------------
\subsection{Reconnection sites}
%--------------------------------------------------------------------
%--------------------------------------------------------------------

%
\begin{figure}
	\centering
	\includegraphics[width=0.45\textwidth]{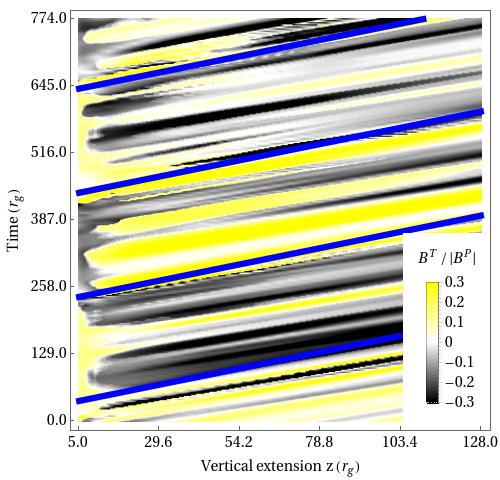}
	\vspace{-6pt}
	\caption{Time evolution of the toroidal dominance $B^T/\left|B^P\right|$ (in a cone enclosing an angle of $\theta=7.5^\circ$, averaged) around the central axis in the C-L4-L2-F model. The \textit{stripes} of alternate polarity propagate through the magnetosphere with the speed of light (slope of the blue lines) carrying opposing polarity of $B^T$. These structures define what we denominate \textit{striped jet}. A new jet stripe is launched with every new accreted flux-tube after a typical time $t_{\rm acc}$ (separation of the blue lines).}
	\label{fig:StripedJet}
\end{figure}
The development of regions with relatively sharp transitions of polarity in the magnetic field (current sheets) is enhanced in 3D compared to axisymmetry. 3D dynamics yield both, a growth of the surface where the magnetic field changes its polarity, and small-scale structures where the magnetic field folds into itself. Thus, they enhance the prospects for (physical) resistive reconnection compared to axisymmetric models. In our numerical method, both of these effects result into (numerical) dissipation of the magnetic field \citep[see, e.g.][for a deep discussion on the similarities of numerical and physical resistive effects]{Rembiasz_2017ApJS}, in qualitative agreement with recently presented simulations by \citet{Bromberg_2019ApJ...884...39,Davelaar_2019arXiv191013370}. A relevant difference between the 2D models of \cite{Parfrey2015} and ours originates from the geometry and surface area of the current sheets between consecutive loops of alternate polarity. When loops plunge into the BH, the shape their common interface is paraboloidal and axial-symmetric, with a surface $S_{\rm 2D}$. In contrast, in 3D it is a wound up paraboloid whose projection on the equatorial plane resembles a helical structure (Figure~\ref{fig:counter_equatorial}). The surface of these wound up structures is (roughly) $S_{\rm 3D}\sim NS_{\rm 2D}$, where $N\sim \Omega_{\rm BH}/(2\Omega_{\rm ISCO})$, and $\Omega_{\rm ISCO}=(a\pm r_{c,{\rm ISCO}}^{3/2}/\sqrt{M})^{-1}$ is the rotational frequency at the ISCO \eqref{eq:VelocityOuter}. For counter-rotating AD models, $N\simeq 4$ ($N\simeq 1$ in the co-rotating case).

The quasy-concentric layers where the magnetic field alternates polarity are potentially well suited to develop \textit{ideal} resistive tearing modes on parallel current layers. The term \textit{ideal} was introduced by \cite{Pucci_Velli:2014}, who showed that current sheets with appropriate thickness $a=S^{-1/3}L$, are unstable against a tearing mode growing on an Alfv\'en (\textit{ideal}) timescale in classical resistive MHD (here $L$ is a characteristic macroscopic length of the current sheet, and $S$ is the Lundquist number; $S\gg 1$ in astrophysical environments, e.g. $S\sim 10^{12}$ in the solar corona). This result has been later confirmed numerically in (special) relativistic resistive MHD \citep{DelZanna_etal:2016,Miranda_2018MNRAS.476.3837}. An extension of this result to multiple-layered systems in resistive relativistic MHD suggests that the growth rate of the tearing mode instability can be even faster than for single current layers \citep{Baty_etal_2013}, even explosive \citep{Baty_2017,Miranda:2018}. Producing the development of these violent reconnection events with 3D global numerical simulations is, so far, not possible because of the extreme computational resources such problem demands.
\footnote{In order to properly resolve the non-linear growth of the fastest-growing tearing mode in the explosive phase, one needs, at least 100 zones per current sheet width \citep[$\Delta x \gtrsim a/100$; ][]{Miranda:2018}. In our models, the typical length of the current sheets is $L\sim 1-10r_g$. Hence we estimate a typical number of numerical zones per dimension of $~[10^6 - 10^7] \times (S/10^{12})^{1/3}$ for producing explosive reconnection events resulting from the relativistic ideal tearing mode instability.} However, it remains to be seen that explosive reconnection may be produced if the simplifications introduced in local numerical simulations are removed. Among the most striking differences between the idealized setup of local numerical models and our global models we single out two. First, the multidimensional geometry of the non-perfectly parallel layers of alternate polarity. Second, the non-stationary dynamics of our current sheets. As such, they are advected, bent, and distorted by local dynamics, i.e. they are strongly perturbed with respect to the optimal configurations for the growth of the ideal resistive tearing mode instability.

Plasmoids emerge following the current sheet that sets limits to the jet launching region around the rotational axis of the BH during the magnetospheric rearrangement between subsequent efficiency peaks (Figure~\ref{fig:counter_bpbr}\textbf{\textsf{b}}). While such instabilities are likely to be sensitive to the imposed accretion model and numerical resolution, the fact that the presented simulations have peak and average efficiencies which are comparable to those obtained by other authors \citep{Parfrey2015} is remarkable. The average efficiencies of our models (see Figure~\ref{fig:Counter_Average} for counter-rotating disc models) deviate by $\sim 15\%$ from the ones derived by \citet{Parfrey2015}, confirming that the accretion of zero net magnetic flux onto fast spinning BHs may also produce intermittent and efficient outflows in 3D. We stress again that our 3D models are resolution limited. Thus, the exact values of the BZ efficiency may change (likely within less than a factor of a few) if larger numerical resolutions (smaller dissipation) were employed.

\citet{Ball2019} show numerically the importance of X-points in reconnection layers for the relativistic acceleration of charged particles. \citet{Guo2019} conclude that such acceleration points are subdominant to the Fermi-type process in reconnection layers, while \cite{Petropoulou2019} recently confirmed the role of X-points in elongated current sheets. Both \citet{Ball2019} and \citet{Petropoulou2019} stress the important role of non-ideal (violating the force-free condition~\ref{eq:FFCondI}) electric fields \citep[and, more recently,][]{Kilian2020}. Such violations are numerically corrected by the our GRFFE scheme (discussed in section~\ref{sec:force-free}), but are likely to occur in regions where the magnetic field rearranges through the formation of small-scale structures that eventually reach the grid scale. Since we are resolution limited, we cannot follow the process of turbulent dissipation of these structures below the grid scale and the numerical scheme reacts by restructuring the electric fields in regions where condition \eqref{eq:FFCondI} needs to be numerically enforced. Customary, this numerical process is considered as an indication of a (potentially turbulent) magnetic field reconnection and is identified with the development of plasmoids along current sheets. Figure~\ref{fig:counter_bpbr} shows a well developed chain of plasmoids at the current sheet flanking the outflow formation region, which are, thus, potential locations of strong particle acceleration and hard X-ray flares \citep{Beloborodov2017,Sironi2019}. As mentioned above, the topology of the magnetic field in the previous current sheet is not axisymmetric, but helicoidal (see the 3D topology of the Poynting flux in Figure~\ref{fig:3DCounter}). Thus, the 2D poloidal maps displayed in Figure~\ref{fig:counter_bpbr} do not show all the small-scale plasmoids developing in that current sheet. This main site for reconnection outside of the AD (turbulent reconnection very likely takes place inside the AD, but this is not included in our simplified model) may host particle acceleration and, hence, time-dependent high-energy processes. As \cite{Yuan2019,Yuan2019a} point out, these reconnection sites relatively close to supermassive BHs in Seyfert galaxies, may produce the hard X-rays responsible for the observed fluorescent emission.

The opposite polarity of subsequently accreted magnetic flux tubes triggers the launching of transient jets with opposite polarity of both the toroidal and the poloidal field. Globally, the magnetic topology of the polar outflows resembles that of an \textit{striped jet}. Jet stripes \citep[cf.][]{Drenkhahn2002,Drenkhahn2002a,levinson2016,Giannios2019} propagate through the magnetosphere at nearly the speed of light (Figure~\ref{fig:StripedJet}) and with a typical stripe length $l_{\rm S}\simeq c\Delta t_{\rm acc} \simeq l c/v_0$. Striped jets provide additional locations for energy dissipation through reconnection at layers of polarity changes, driving both the jet bulk acceleration and particle energization, though likely at larger scales than we have considered here \citep[][and references therein]{Giannios2019}.

%--------------------------------------------------------------------
%--------------------------------------------------------------------
\subsection{Ideal loop efficiency}
\label{sec:loopefficiency}
%--------------------------------------------------------------------
%--------------------------------------------------------------------

\citet{Uzdensky2008} propose that the formation of loop structures with sizes significantly larger than the AD height is possible by reconnection in the disc corona, supporting our (simplified) setup. Our analysis suggests a broad range of loop areas around $l\times h\approx 7.4 \,r_g^2$ for an optimal process efficiency during the accretion of magnetic loops from a counter-rotating AD (see Figure~\ref{fig:Counter_Average}). We also find indications of a significant decrease of the process efficiency for both very small ($\simeq l\times h\approx 1.7\, r_g^2$) and large ($\simeq l\times h\approx 32.1\, r_g^2$) loop cross-sectional areas.

The accretion of magnetic flux tubes from co-rotating AD models extracts energy from the central BH much less efficiently. The time-evolution of magnetic fields recurrently establishes configurations in which all field lines emerging from the BH connect to the AD \citep[as in the equilibrium solutions of, e.g.][]{Uzdensky2005, Mahlmann2018,Yuan2019a,Yuan2019}. The resulting transport of angular momentum from the BH to the AD combined with the (artificial) magneto-rotational energy extraction from our (simple) disc model (appendix~\ref{sec:faraday_disc}) gradually distorts the UAL of the disc throughout the simulations. However, besides the negative feedback of these effects on the transport of energy from the BH to infinity, when field lines open up in the polar regions (cf. Figure~\ref{fig:counter_bpbrco}), peak efficiencies of $\epsilon\approx 0.2$ are reached. 

\begin{figure*}
	\centering
	\includegraphics[width=0.925\textwidth]{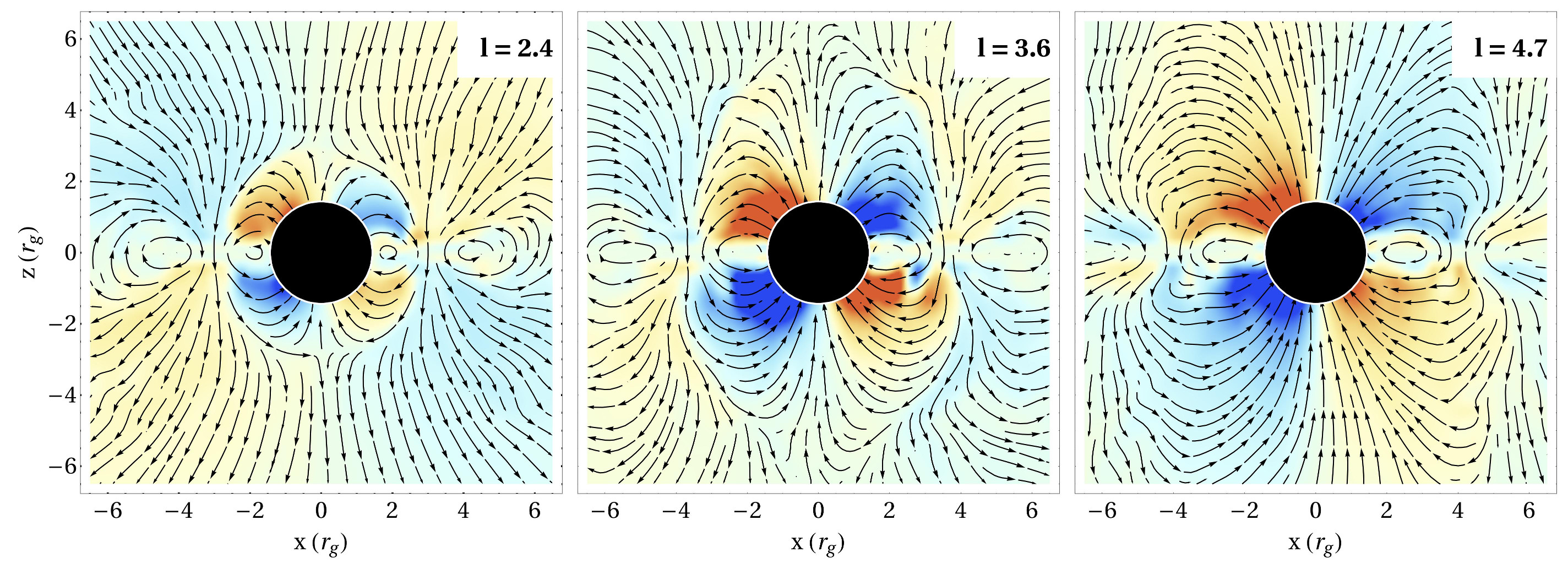}
	\vspace{-6pt} 
	\caption{Toroidal field $B^T$ (\textit{blue} negative, \textit{red} positive) for selected tests of a co-rotating AD with ($\alpha=2$, $r_1=r_{\rm ISCO}$, $v^r=0$) and a completely force-free plunging region. The loop length $l$ is varied. Poloidal magnetic field lines are overlaid. The BH and AD rotation are set up as in our models (see section~\ref{sec:numerical_setup}). These ancillary models are a GR extension to the ones of \citet{Yuan2019} with the notable difference that \citet{Yuan2019} do prescribe boundary conditions on the electromagnetic fields at the BH horizon (employing a membrane description of the BH). The presented snapshots are taken at $\sim 9$ revolutions of the BH ($\sim 180 r_g$) after initialization and correspond roughly to the variability timescales $\Delta t_{\rm acc}$ present in our slowly accreting models (series A). We observe the progressive twisting of the magnetic field close to the BH, manifesting itself as a pair of growing lobes with a \textit{butterfly} shape in the poloidal plane.}
	\label{fig:strength_model}
\end{figure*}

\citet{Yuan2019a,Yuan2019} relate the emergence of open field lines in BH/AD systems to the ratio between the magnetic flux of the inner loop to the outer one. In their toy-model, consisting of two flux tubes of length $l\approx 2.5r_g$ (appendix~\ref{sec:yuan_benchmark}), they use a current similar to ours (\ref{eq:InitCurrent}) but adding the possibility that the current decays radially as $r^{-(\alpha+1)}$. In practice, they mimic the effect of frame dragging (or rigid rotation) by imposing an appropriate surface resistivity to a central disc-shaped membrane in a special relativistic simulation. These BH/AD toy-models show a strong dependency of the field line topology on the decay parameter $\alpha$. The stronger the interior flux tube compared to the outer one, the more (and the faster) inclined the field lines emerge until they eventually open up. In other words, if stronger magnetic fields connect the tip of the AD to the BH in the plunging region, it is more likely to open up the magnetic field lines or to develop vertical, ordered magnetic structures. In order to validate our own results, we have reproduced the numerical setup suggested by \citet{Yuan2019a} in appendix~\ref{sec:yuan_benchmark}, and elaborate on it employing our GRFFE code. For that, we have run a set of ancillary models in which we use an AD setup, which combines the essential magnetospheric structure of \citet{Yuan2019a} with our co-rotating models (Figure~\ref{fig:strength_model}). Especially, we have focused on the establishment of an equilibrium of loops for $\alpha=2$ with different loop-length $l$. These loop lengths have been chosen to bracket the critical loop length $l_{\rm crit}=3.2r_g$ obtained by \citet{Parfrey2015}. We find that a radially outwards Poynting flux occurs even for $l=2.4\,r_g<l_{\rm crit}$ for these ancillary models. However, the lobes growing around the central BH have a finite size and do not efficiently connect to infinity (except, perhaps, along a bundle around the rotational axis of the system with a tiny radius). Thus, models with such an small loop width (below $l_{\rm crit}$) are not expected to produce outflows. We observe the growth of a pair of lobes around the BH with a \textit{butterfly} shape in the poloidal plane in Figure~\ref{fig:strength_model}. These lobes become larger with increasing $l$, optimizing the prospects for the emanation of Poyting flux to infinity. The ancillary models shown in Figure~\ref{fig:strength_model} cross-validate our results in several ways. First, we find that the BZ process is activated even for loop lengths below $l_{\rm crit}$. Second, they qualitatively reproduce the results of \cite{Yuan2019}, employing a resolution similar to the models of this paper. Third, larger loop lengths yield poloidal fields, which make a smaller angle to the vertical direction, hence improving the available efficiency of the BZ process. Finally, we find that with a different current distribution on the AD (with different radial dependence, see appendix~\ref{sec:yuan_benchmark}), the previous conclusions still hold. The combination of the AD setup in \citet{Yuan2019} with the full GR capacities of our method is well suited to analyse the influence of the rotation of an idealised disc on the activation of the BZ mechanism. This is because the effects of rotation are gauged by both, a suitable choice of the electric fields on the equatorial membrane mimicking the AD, and frame-dragging of the spacetime itself. By construction, the latter is not included in \citet{Yuan2019}. Furthermore, only fieldlines connecting to the central membrane can contribute to the magnetospheric energy flows in their default model of no AD rotation. While such idealised models provide a clean picture for the magnetospheric dynamics induced by BH/AD differential rotation, the setup from \citet{Yuan2019} cannot straightforwardly be extended to account for the full \textit{accretion} dynamics (i.e. for the radial displacement of the magnetic field lines). This is one of the most distinctive elements of our models compared to the ancillary setups we have considered above.

With a shorter accretion time $\Delta t_{\rm acc}$ (i.e. larger accretion speed $v_0$), field lines connecting the BH to the AD have less time to be twisted by differential rotation. In case of our co-rotating models, the central object completes $\sim 1.4$ revolutions per each rotation of the tip of the AD at $r_{\rm ISCO}$. In case of the counter-rotating models, the BH spins $\sim 7.8$ times in the opposite direction during one turn of the field lines located at $r_{\rm ISCO}$. Flux tubes which accrete without being sufficiently twisted by differential rotation may fail to develop sufficiently vertically elongated poloidal magnetic field lines. This vertical structure of the magnetic field (optimally connecting the BH to infinity) is required to drive an outflow (see Figures~\ref{fig:counter_bpbr} and~\ref{fig:StripedJet}) under ideal conditions for operation of the BZ process (appendix~\ref{sec:bz_signatures}). However, closed magnetic field lines linking the BH and the AD may transport energy and angular momentum between them, as in case of co-rotating AD models.\footnote{In our models the feedback on the BH of the transport of energy and angular momentum from the AD is not included, since we do not feed the space-time evolution with the dynamics of the magnetic field (Cowling approximation; see section~\ref{sec:simulations}). Likewise, since the \textit{velocity} in the AD is imposed numerically, instead of being the result of a self-consistent MHD calculation, the transport of energy and angular momentum from the BH to the AD does not result into a braking or speeding up of the latter.} The dynamics of these closed magnetic field lines is very important to set the efficiency of the BZ process, and we observe a contrasting behaviour in co- and counter-rotating AD models. We find that some of the closed field lines connecting the BH to the AD experience a premature detachment from the AD due to 3D instabilities. These (kink-like) instabilities manifest in some models as, e.g. the \textit{fall-down} of an incipient \textit{magnetic tower} \citep[i.e. a vertical thick flux tube along the symmetry axis threaded by helicoidal magnetic field lines; the basis of the magnetic tower emerges as helical patterns in the movie associated to Figure~\ref{fig:counter_equatorial}; see also][]{Lynden-Bell_1996MNRAS}. The development of kinks due to non-axisymmetric effects has also been noticed by \citet{Yuan2019}, who estimated that the timescale for the growth of these kinks in the outflow is
\begin{equation}
t_{\rm kink}\simeq (\hat{h}/\hat{k} + 1)^2 r_{\textsc{lc}},
\label{eq:tkink}
\end{equation}
where $r_{\textsc{lc}}\simeq 1/\Omega_{\rm BH}\simeq 3.2\,M$ is an estimation of the radius of the light cylinder. $\hat{h}=z/r_{\textsc{lc}}$ and $\hat{k}=z/(r_c-r_{\textsc{lc}})$ are the vertical height of the outflow in units of $r_{\textsc{lc}}$ and a constant characterizing the opening angle of the outflow, $\theta_{\rm out}\sim \arctan{(1/\hat{k})}$, respectively. 
Kink instabilities may grow in a steady, expanding, collimated outflow if their characteristic growth time \eqref{eq:tkink} is shorter than the expansion timescale of the flow,
\begin{equation}
t_{\rm flow}\simeq \hat{h} \sqrt{1/\hat{k}^2+1} \, r_{\textsc{lc}},
\label{eq:tflow}
\end{equation}
i.e. if $t_{\rm kink}/t_{\rm flow}<1$ \citep{Yuan2019}. In our models, the outflow opening angle is not easy to compute. One possibility we have adopted is to evaluate the angular location, measured from the vertical axis, where the (radial) Poynting flux changes sign on a spherical surface with radius $r_{\rm P}$. Certainly, the opening angle is a function of the radial distance to the BH. Hence, to quantify our results, we measure the opening angle relatively close to the BH, namely, at $r_{\rm P}=16\,M$. We pick this value because we observe that kinks in the magnetic tower can already develop at smaller values of $r$. We cautiously point out that $r_{\rm P}/r_{\textsc{lc}}\gg 1$ to ensure that the approximations employed to derive \eqref{eq:tkink} and \eqref{eq:tflow} hold. We note, however, that in our case $r_{\rm P}/r_{\textsc{lc}}\gtrsim 5$ and, more importantly, the outflow is not stationary. Hence, 
the estimates \eqref{eq:tkink} and \eqref{eq:tflow} are only crude approximations.

\begin{figure}
	\centering
	\vspace{-12pt}
	\includegraphics[width=0.48\textwidth]{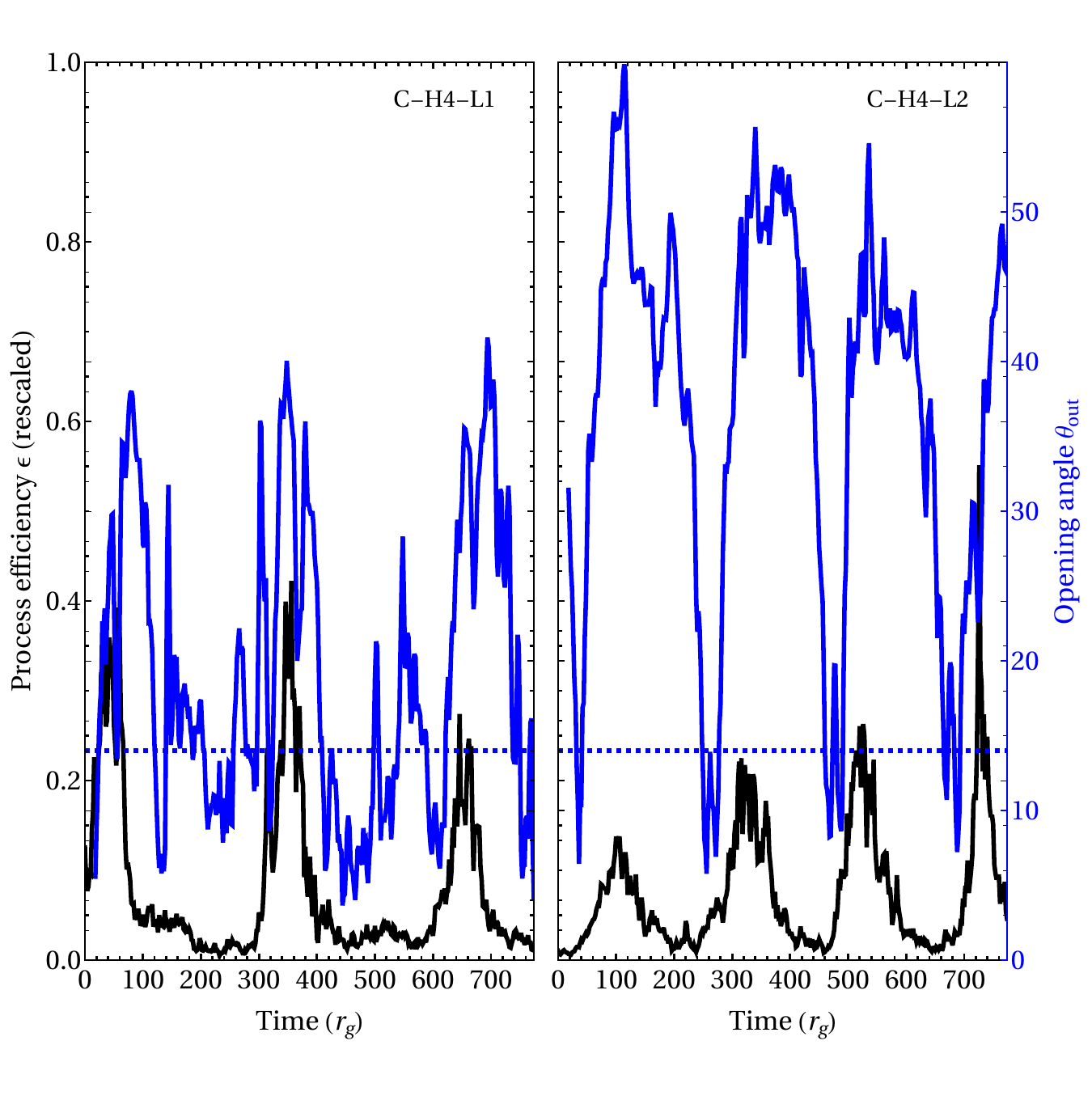}
	\vspace{-26pt} 
	\caption{Evolution of the process efficiency (black curves, rescaled by a factor of five) and the opening angle $\theta_{\rm out}$ (blue solid lines) for selected models of series A. The horizontal blue dashed line marks the value $\theta_{\rm kink}\approx 14^\circ$. C-H4-L1 (\textit{left}) shows smaller outflow opening angles (on average) than C-H4-L2 (\textit{right}). For C-H4-L1, values $\theta_{\rm out}<\theta_{\rm kink}$ are observed in clear correlation with periods of low BZ efficiency, during timescales comparable to $\Delta t_{\rm acc}$. In case of the efficiently working reference model C-H4-L2 (\textit{right}), the flow develops significant opening angles of $\theta_{\rm out}\sim 50^\circ$ and extending before and after any of the high-efficiency BZ peaks. Short episodes of $\theta_{\rm out}\sim\theta_{\rm kink}$ are transient events and happen in association with the change of polarity throughout the magnetosphere (see section~\ref{sec:field_structure}).}
	\label{fig:OpeningAngle}
\end{figure}

The values of $\theta_{\rm out}$ measured at $r=r_{\rm P}$ are extremely time dependent. They change from nearly zero to $\lesssim 60^\circ$ on timescales that are shorter than $\Delta t_{\rm acc}$ (see Figures~\ref{fig:OpeningAngle} and ~\ref{fig:seriesb_energy}). Kink instabilities may only set in when the outflow opening angle is sufficiently small, i.e. when $\theta_{\rm out}\le \theta_{\rm kink}\simeq 14^\circ$; corresponding to $\hat{k}\simeq 4$. Values of the outflow opening angle smaller than $\theta_{\rm kink}$ happen before and after the peaks of efficient BZ energy extraction. When $\theta_{\rm out}\sim \theta_{\rm kink}$, the ratio $t_{\rm kink}/T_{\rm ISCO}\simeq 1$, i.e. the kink growth timescale roughly coincides with the orbital period at the ISCO ($T_{\rm ISCO}=2\pi/|\Omega_{K}(r_{\rm ISCO})|$). 

For models with short loop length (see Figure~\ref{fig:OpeningAngle}) or faster accretion speeds (see Figure~\ref{fig:seriesb_energy}), i.e. shorter accretion times, the opening angle tends to be smaller than for wider loops or smaller values of $v_0$ (with extended periods of $\theta_{\rm out}\lesssim\theta_{\rm kink}$).
 In our simple estimate of the typical accretion speed in a Shakura-Sunyaev disc \eqref{eq:v_0max}, both cases are disjoint. Larger values of $l$ are connected to larger values of the disc half-thickness in our model and, hence to larger values of $h/r$, which make the accretion speed ($v_0\simeq \alpha_{\textsc{ss}}(h/r)^2 v_{\rm K}$) grow. Conversely, smaller accretion speeds are linked to a smaller disc half-thickness, which in our model setup imply smaller values of $l$.
	
Analysing the time evolution of the periodogram of the vertical magnetic field on a slim ring of radius $2M$, we find the growth of modes with wavelength comparable or a few times shorter than the length of the ISCO orbit ($L_{\rm ISCO}\simeq 55\,M$) during the periods of efficient BZ energy extraction. These wavelengths correspond to time-scales a few times shorter than $T_{\rm ISCO}$. Conversely, in between of high efficiency peaks, modes with shorter wavelengths appear. The length and timescales of these structures is correlated with the loop length: Larger values of $l$ develop longer wavelengths and shorter timescales. The alternation of shorter and longer dominant modes in the equatorial plane corresponds to the formation of spiral structures in the plunging region during the luminosity bursts and their disappearance in periods of low efficiency $\epsilon$. While we have observed such indicators of a loss in equatorial and axial symmetry for different mesh resolutions (and distributions), detailed field dynamics in the plunging and jet regions will be probed with high resolution simulations in the near future.

The counter-rotating reference model C-H4-L2 develops dynamics comparable to the axisymmetric model of \cite{Parfrey2015}, hence minimising the role of 3D instabilities in the efficiency of the BZ process. In contrast, the counter-rotating models C-H4-L1 and C-H4-L2-02 show alternations between very efficient extraction cycles due to the accretion of one AD magnetic flux tube and a subsequent, significantly less efficient period. In these models, the hairpin structures developing in the plunging region do not sufficiently extend vertically and only drive a partial change in the polarity of the magnetosphere in the vicinity of the BH. The energy extraction by the BZ process is suppressed for these \textit{insufficiently stretched} magnetic structures. We note that in axial symmetry, the differential rotation between the disc and the BH inevitably yields to increasing the toroidal twist and, eventually, to open up the field lines to infinity \citep{Uzdensky_2002}. This is a consequence of the relativistic Ferraro's Law of isorotation \citep[c.f.][]{Yuan2019}, which states that the angular velocity along a field line must be constant. Without imposing axial symmetry, this is not necessarily the case. Thus, the prospects to open up magnetic field lines leading to an efficient energy extraction are smaller in 3D than in 2D.

In the case of co-rotating BH/AD models, the decay of the magnetic field strength in the accretion funnel due to 3D instabilities (see above), may greatly impair the development of strong BZ type outflows. Figure~\ref{fig:counter_bpbrco} and section~\ref{sec:field_structure} identify this interplay between the tendency to connect the BH and the AD by closed magnetic field lines in a (short-term) quasi-equilibrium structure \citep[akin to the magnetostatic configurations of][]{Uzdensky2005}, and short periods in which open (or insufficiently stretched) field lines drive relatively low-power outflows.

 Our numerical models, backed up by the ancillary simulations employed to compare to \cite{Yuan2019a}, suggest that it is necessary to allow for several tens of rotational periods of the central object in order to build up inclined structures of twisted magnetic field lines (see Figures~\ref{fig:strength_model} and~\ref{fig:membrane_rotator}). Thus, there exists an additional relevant timescale to set the efficiency of the BZ process in our setup. This is the time required to sufficiently and uniformly twist the magnetic loops connecting the BH to the AD.
 We can estimate this timescale as
 \begin{equation}
t_{\rm w}= \frac{2\pi}{0.5\Omega_{\rm BH}-\Omega_{\rm ISCO}},
 \end{equation}
where we compare the rotational frequency at the ISCO to half the rotational frequency of the BH (as this is the optimal value of the field-line angular velocity to drive an efficient BZ process, see appendix~\ref{sec:field line_velocity}). Our models have values of $t_{\rm w}\simeq 91\,M$ and $t_{\rm w}\simeq 32\,M$ for co- and counter-rotating discs, respectively.
 
A thorough understanding of the role of the loop size and the loop magnetic field strength for the development of dissipative regions (by either kink instabilities or by reconnection processes) or relativistic outflows (BZ jets) will be studied in a subsequent work.

%--------------------------------------------------------------------
%--------------------------------------------------------------------
\subsection{Variability timescales of the BZ luminosity}
\label{sec:variability_timescales}
%--------------------------------------------------------------------
%--------------------------------------------------------------------
% 
\begin{table}
	\centering
	\caption{Timescales. The columns represent from left to right, the identifier, the values of $h$ and $l$ in units of $r_g$, the ratio of the accretion timescale $\Delta t_{\rm acc}\simeq l/v_0$ to the Keplerian orbital period at the ISCO, $T_{\rm ISCO}=2\pi/|\Omega_{\rm ISCO}|$, and the orbital period at the ISCO (in units of the BH mass).}
\label{tab:models}
	{\renewcommand{\arraystretch}{1.5}
		\begin{tabular}{ccccc}
			\hline
			Model & $h$ & $l$ & $\Delta t_{\rm acc}/T_{\rm ISCO}$ &$T_{\rm ISCO}$\\*[0pt]\hline
			C-H4-L2-005 & 4.0 & 2.0 & 3.68& 158\\
			C-H4-L2-01 & 4.0 & 2.0 &1.84 & 158 \\
			C-H4-L2-02 & 4.0 & 2.0 & 0.92 & 158 \\
			C-H4-L2-04 & 4.0 & 2.0 & 0.46 & 158 \\\hline			
			R-H1-L2 & 1.0 & 2.0 & 10.3 &32.4\\
			R-H1-L3 & 1.0 & 3.0 & 15.5 &32.4\\
			R-H1-L4 & 1.0 & 4.0 & 20.6 &32.4\\\hline
	\end{tabular}}
\end{table}
\begin{figure*}
	\centering
	\includegraphics[width=0.98\textwidth]{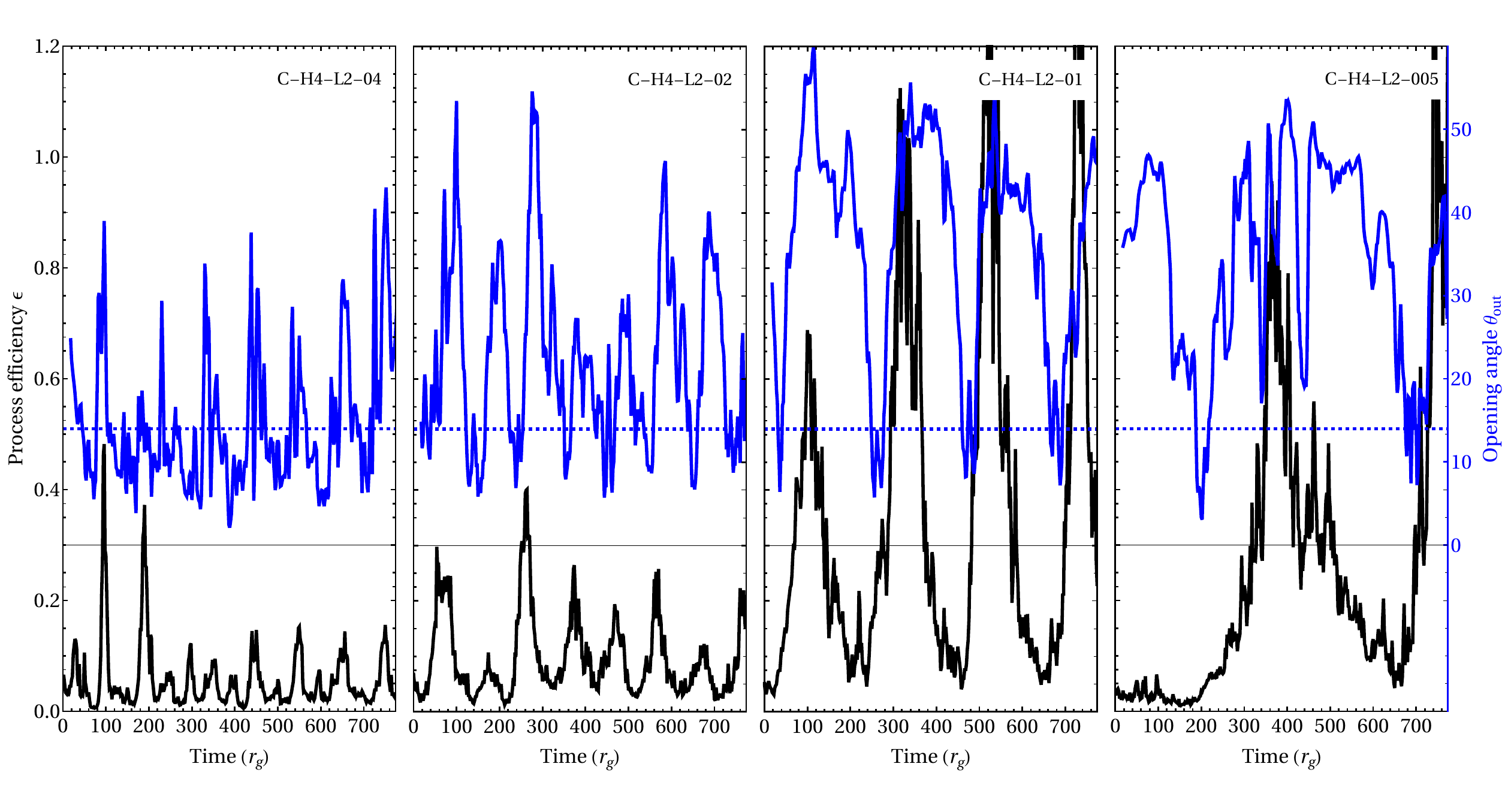} 
	\vspace{-18pt}
	\caption{Instantaneous efficiency $\epsilon$ (eq.~\ref{eq:process_efficiency})of the BZ process and opening angle $\theta_{\rm out}$ during the evolution of the counter-rotating disc reference model C-H4-L2 with different accretion speeds (series B, cf. Table~\ref{tab:model_initials}). The pulse periodicity changes according to the accretion velocity $v_0=\left[0.005,0.01,0.02,0.04\right]$. We employ the same colors than in Figure~\ref{fig:OpeningAngle}. Our results suggest a roughly linear dependence of the time-averaged outflow opening angle and the accretion speed for the models of the figure $\langle \theta_{\rm out} \rangle \simeq 0.7 - 11 \times v_0/c \,$.
}
	\label{fig:seriesb_energy}
\end{figure*}

The presented simulations are conducted in a system of units scaling with the BH mass $M_{\rm BH}$. Especially, timescales are directly proportional to the mass of the central object. For relatively small values of the accretion speed (see below), the accretion timescale $\Delta t_{\rm acc}\simeq l / v_0$ determines the overall duration of a single luminosity burst or pulse $T_{\rm Pulse}$:
\begin{align}
T_{\rm Pulse}=1.37\times\left(\frac{c}{v_0}\right)\left(\frac{l}{r_g}\right)\left(\frac{M_{\rm BH}}{M_9}\right)\:\text{hrs}.
\label{eq:Tpulse}
\end{align}
Here, we employ $M_9=10^9M_\odot$. For a loop length of $l=2r_g$, $v_0=0.01c$ (corresponding, e.g. to our reference counter-rotating model C-H2-L2) and a BH of $M_{\rm BH}=6.5M_9$ this yields an approximate pulse duration of 74 days. We would like to stress that the accretion velocity, $v_0=0.01$ for the principal set of simulations (series A), has been chosen for numerical convenience and approximate comparability to the results of \citet{Parfrey2015}. On the basis of a crude estimation, we have argued that we expect finding typical values $v_0\lesssim 0.02 c$ \eqref{eq:v_0max}. Hence,
series B (see Table~\ref{tab:model_initials}) assembles variations of the reference model C-H4-L2 ($v_0=0.01$) with different accretion speeds $v_0=\left[0.005,0.02,0.04\right]$. The pulse duration $T_{\rm Pulse}$ scales well with the chosen accretion speed, as we show in Figure~\ref{fig:seriesb_energy}. However, the faster accretion speed counteracts the necessary spinning up of field lines connecting the BH to the AD and may cause insufficiently stretched magnetic structures (see previous section) in between energy peaks. Thus, the pulse duration estimated by \eqref{eq:Tpulse} should be taken with care, due to the non-linear nature of our results, and due to the fact that the process efficiency does not linearly depend on the loop length (Section~\ref{sec:Counter-rotatingAD_efficiency}). It could be that in case of fast accretion speeds a larger efficiency may be obtained for larger loop lengths (at the cost of lengthening the duration of pulses from equation~\ref{eq:Tpulse}). Understanding in detail how the output power and variability timescale changes with $v_0$ and $l$ requires considering different models of ADs and initialisation of the magnetic flux tubes inside them, something beyond the scope of this paper. Forcing the estimate in \eqref{eq:Tpulse} to its limit of validity ($v_0\simeq 0.04$, $l\simeq 1r_g$), one ends up with $T_{\rm Pulse}\simeq 9\,$days for a BH mass like that of M87. Such timescales are somewhat longer than the shortest variability timescales observed in the TeV radiation of M87 \citep{Acciari2009,Aharonian2003} and part of the data from the radio galaxy IC310 \citep{Aleksic2014}. However, the shortest variability timescales in these AGNs correspond to extreme flaring events, which may require special conditions to develop. 

Besides the most evident variability timescales, roughly corresponding to $\Delta t_{\rm acc}$, smaller variability timescales are present within each pulse. A spectral analysis of the data in, e.g. Figure~\ref{fig:counter_energy} (assembling models with $v_0=0.01c$) reveals that there is significant power at timescales of up to $\sim 4$ times smaller than $\Delta t_{\rm acc}$. In our counter-rotating models the period at the ISCO is $T_{\rm ISCO}=158\,r_g\sim \Delta t_{\rm acc}/2$ (Table~\ref{tab:models}), hence, we identify the spectral power observed at frequencies $\simeq 2/\Delta t_{\rm acc}$ with the dynamics of the loops as they are released from the ISCO. Certainly, the ratio $\Delta t_{\rm acc}/T_{\rm ISCO}$ depends on the imposed accretion speed and we foresee that variability timescales associated to the location of the ISCO are longer than $\Delta t_{\rm acc}$ for $v_0\gtrsim 0.02c$ (see Tab.\,\ref{tab:models}).
The amplitude of the variability at timescales below $\sim \Delta t_{\rm acc}/2$ is nearly two orders of magnitude smaller than that corresponding to $\Delta t_{\rm acc}$. Both, the amplitude of these variations and, consistently, their spectral power depend on the numerical resolution since they are linked to resistive effects. Running the complete set of models at higher resolution than we have done so far demands extremely large computational resources. Besides, it is not justified to employ only a force-free model that does not account for the mass-loading of the magnetosphere and other non-ideal MHD effects. Thus, we cannot robustly assess the variability at timescales below $\sim \Delta t_{\rm acc}/4$ with our models.
%

%--------------------------------------------------------------------
%--------------------------------------------------------------------
\section{Conclusions}
\label{sec:conclusions}
%--------------------------------------------------------------------
%--------------------------------------------------------------------

We have conducted a set of simulations of a simplified model advecting tubes of zero net magnetic flux in a thin \textit{ad hoc} AD towards a rapidly spinning central BH. In total, we have probed 16
 different BH/AD parameter setups plus 3 ancillary models for benchmarking. Our results show that powerful, intermittent outflows, driven by the BZ mechanism can form in 3D from magnetic structures having scales a few times larger than the AD height. Admittedly, our setup is a simplification of an actual AD, which does not allow for any torque from the BH onto the disc. Our AD toy-model is better suited for counter-rotating than for co-rotating ADs. In the latter case, there is a flux of energy resulting from the fact that our simplified setup makes the AD behave as a Faraday disc which is both accelerated and loaded with a current (appendix~\ref{sec:faraday_disc}). In spite of this artificial effect, our prograde AD models are still adequate to understand the energy flows at the horizon and in the immediate vicinity of the central BH for values of $a^*\lesssim 0.9$.

The average efficiency of the BZ process is very significant, $\bar{\epsilon}\sim 0.4$ for counter-rotating ADs, and $\bar{\epsilon}\sim 0.1$ for co-rotating ADs. Episodes of efficient energy extraction are linked to ordered magnetic fields structured by a dipolar component in the plunging region, as well as field lines emerging from the BH and extending to a significant distance. This effectively creates ideal conditions comparable to those of \citet{Blandford1977} for a short period of time. Even without imposing large-scale magnetic flux structures, our models develop an outgoing Poynting flux at the BH horizon in broad accordance with the BZ mechanism. At the same time, 3D \textit{dynamical} evolution also triggers complicated field structures and field reversals, which cannot be described by the equations of \textit{stationary}, \textit{axisymmetric} GRFFE. In these regions, the efficiency of the BZ process is reduced or breaks down completely.

We have made a study over a range and combinations of parameters defining the accretion disc structure. They confirm and extend the work by \citet{Parfrey2015}, especially in characterizing the Poynting outflow efficiency during periods of energy extraction.
In line with \cite{Parfrey2015} our results also suggest that jets may be quenched in prograde accretion flows if there is no large scale magnetic field threading the BH/AD system. Since our model setup is not exactly the same, we find a slightly smaller value of the critical loop length than \citet{Parfrey2015} did. A more careful modelling of the conditions in different environments, where there may exist prograde accretion flows around rapidly spinning BHs, is needed to more robustly assess the generality of a jet quenching mechanism that depends upon the ability of the turbulent AD to create magnetic flux tubes with sufficiently large sizes. Some of these systems are found, e.g. in Seyfert galaxies \citep{Risaliti_2013Natur.494..449}, intrinsically X-ray weak intermediate-mass BHs \citep[e.g. in PL~1811][]{Dong_2012}, where the small variability timescales indicate that the X-ray source is rather compact, or in X-ray binaries during the soft state \citep[e.g.][]{Plant_2014MNRAS.442.1767}.

Our models are resolution limited, since we aimed to run a relative large number of full-fledged 3D simulations. This means that resistive effects may be (numerically) overestimated. We have, however, benchmarked our results agains the (simpler) setup of \cite{Yuan2019}, employing a numerical resolution similar to the one employed in the rest of the models of this paper. We have found a remarkably good qualitative agreement with the results of these authors. In particular, ancillary models set up to reproduce a different radial distribution of the current in the AD \citep[akin to that of][]{Yuan2019}, cross-validate our result that efficiency is tightly linked to the loop length in prograde ADs. This comparison also serves for the purpose of assessing that our findings are not exclusively valid for the (simple) current setup employed here.

Future studies demand higher numerical resolution in a broad wedge around the rotational axis of the BH in order to accurately describe the dynamics of the generated outflows and the expelled plasmoids, as well as their interactions with flows directed along the jet axis. Understanding the detailed structure and the overall dynamics of plasmoids in reconnection regions may allow for links to recent results from first-principle simulations and the interpretation of their radiative imprints \citep[cf.][]{Christie2020}. The viability of our \textit{ad hoc} AD setup should be probed in GRMHD simulations, for example starting off at the flux tube structures considered by \citet{Beckwith2008,Beckwith2009} or the plasmoid formation modeled in 2D GRMHD by \citet{nathanail2020}.

The accretion of zero net magnetic flux structures with opposing polarity triggers quasi-periodic phenomena, most significantly, on a variability timescale given by the accretion time $\Delta t_{\rm acc}\sim l/v_0$, but also on smaller timescales (see below).
For supermassive BHs at the core of AGNs like M87, these time-scales are days-to-months, while for stellar mass BHs in X-ray binaries, they may be as small as $\sim 0.01\,(M_{\rm BH}/10\,M_\odot)$\,s, and for intermediate mass BHs $\sim 100\,(M_{\rm BH}/10^5\,M_\odot)$\,s. Shorter timescales ($\gtrsim \Delta t_{\rm acc}/4\sim 2$ days to two weeks for supermassive BHs) are also reliably observed in our models. They result from the loop dynamics close to the inner edge of the AD and from genuinely 3D (kink-like) instabilities \citep[they are not observable in axisymmetry][]{Parfrey2015}. For supermassive BHs, these variability timescales are comparable to the longest timescales observed, e.g., in the TeV radiation of M87. However, a thorough comparison of our results to observations requires a post-processing step. Namely, computing the emission and radiation transfer from our models to the observer, effectively accounting for lensing effects, caustics, etc. induced during radiation propagation by the BH. This is beyond the scope of this paper, and we will address it elsewhere.

Resistivity is of numerical origin in our models. Hence, it is dependent on the (limited) numerical resolution of 3D models. However, we observe the standard phenomenology induced by the resistive dissipation of the magnetic field when current sheets develop as a result of the gravito-magnetic coupling between the BH and the AD plasma. Specifically:
\begin{itemize}
	\item[i)] Due to the polarity change of the accreted magnetic flux, the outflow developing over large distances along the central axis has a striped structure. In between of each of the jet stripes, additional reconnection sites may form. This structure gives qualitative support to the so-called striped jet model \citep[e.g.][]{Giannios2019}, or models where reconnection is responsible for blazars' gamma-ray emission \citep{Giannios_2013MNRAS.431..355}.
	\item[ii)] Extended current sheets form during the reordering of magnetospheric field polarisation. These current sheets are prone to develop magnetic islands (plasmoids), which are potential locations for relativistic particle acceleration. 
	\item[iii)] Closely connected to the previous current sheets, we identify sheets of alternate polarity in the plunging region and extending vertically above and below the equator. They arise naturally as a result of the differential rotation acting on diametrically opposed ends of the AD loops. Their projection on the equatorial plane resembles an $m=1$ or $m=2$ spiral structure. In planes parallel to the equator, these structures form a set of similarly thick spiral arms of alternate polarity. Such structures bear a topological similarity with sets of parallel current sheets with alternate polarity in which the relativistic \textit{ideal} tearing mode instability may develop. We speculate with the possibility that these locations might develop explosive reconnection events. However, due to the stringent numerical resolution demanded to observe the violent non-linear phase of the relativistic ideal tearing mode instability, it is unlikely that global 3D simulations may unveil it in the very near future.
\end{itemize}

Magnetic reconnection converts magnetic energy into thermal and kinetic energy. The results of our idealised setups support models where reconnection may take place at very different scales, including scales of the order of a few gravitational radii ($\lesssim 10r_g$) in BH/AD systems, independent of the BH mass \citep[e.g.][in the case of X-ray binaries]{Beloborodov_1999ApJ...510L.123}. Reconnection can be the source of the X-ray coronae not only in X-ray binaries, but also in AGNs as well as in intermediate-mass BHs.  Also fast magnetic reconnection between the  magnetic field  lines of the inner disk region and those that are anchored in the black hole has been suggested to produce the radio flares in galactic microquasars such as GRS 1905+105 \citep{GouveiaDalPino_Lazarian_2005A&A...441..845}, as well as in AGNs \citep{GouveiaDalPino2010}.

Although the simulations we have carried out are 3D, the initial setup is axisymmetric. Likely, the azimuthal extension of loops produced in the disc as a result of the magneto-rotational instability may be $\Delta \phi\gtrsim H/r$, as argued by \cite{Parfrey2015}. We obtain, however, that the dynamics of the loops as they detach from the AD are non-axisymmetric. This is due to genuine 3D instabilities in the outflow, where magnetic towers are kinked until they tip over the magnetic flux tubes in the plunging region. Indeed, reconnection combined with the effect of the fall-down of the magnetic towers create extended perturbations in the azimuthal direction. As a result, parts of the same (initially axisymmetric) flux tube at different azimuthal angles interact with the BH asynchronously. Each of these angular sectors contributes to the overall large-scale jet incoherently and, hence they produce the substructure observed in the BZ efficiency plots, within each large-scale outburst. We find that this substructure accounts for variations in the efficiency of the BZ process, one or two orders of magnitude smaller than the (ideal) overall accretion of each of the large-scale concentric flux tubes.

Once the magnetic field becomes strong enough in the vicinity of the BH, it may counteract totally or partially the in-fall, effectively breaking our ad-hoc accretion flow. Thus, future work may go along improving the kinematic approximation used to impose the accretion velocity in the equatorial plane. The coincidence of small opening angles of the outflow (hence, prone to kink instabilities) at some distance from the BH and a non-efficient working of the BZ process at the BH horizon deserves special attention. With the correlation of large opening angles (hence, kink-stable flows) and efficient energy extraction also true in the presented simulations, we conclude that jet launching by gravitomagnetic coupling after all does require a stable magnetic structure extending over several $r_g$. Based on the presented analysis, we expect such structures to form preferentially when the length of the magnetic loops is large (hence, the disc half-thickness is large; $h/r\sim 0.5$), and when the accretion speed (and, likely, the mass accretion rate) is small.

%--------------------------------------------------------------------
%--------------------------------------------------------------------
\section{Acknowledgements}
%--------------------------------------------------------------------
%--------------------------------------------------------------------

We appreciate Kyle Parfrey's constructive review of our manuscript as well as his feedback during the development of this work. We thank Vassilios Mewes, Pablo Cerd\'a-Dur\'an, Serguei Komissarov and Alejandro Torres for feedback (and hospitality) along many steps of this paper as well as fruitful discussions going deep into the details of their respective contributions. JM acknowledges a Ph.D. grant of the \textit{Studienstiftung des Deutschen Volkes}. We acknowledge the support from the grants AYA2015-66899-C2-1-P, PGC2018-095984-B-I00, and PROMETEO-II-2014-069. We acknowledge the partial support of the PHAROS COST Action CA16214 and GWverse COST Action CA16104. The shown numerical simulations have been conducted on infrastructure of the \textit{Red Espa\~{n}ola de Supercomputación} (AECT-2019-1-0004, AECT-2019-3-0017) as well as of the \textit{University of Valencia}. 

%--------------------------------------------------------------------
\bibliographystyle{mnras}
\bibliography{literature}
%--------------------------------------------------------------------

\appendix

%--------------------------------------------------------------------
%--------------------------------------------------------------------
\section{Numerical details}
\label{sec:num_details} 
%--------------------------------------------------------------------
%--------------------------------------------------------------------

%--------------------------------------------------------------------
%--------------------------------------------------------------------
\subsection{The augmented system}
\label{sec:augmented_system} 
%--------------------------------------------------------------------
%--------------------------------------------------------------------

The General Relativistic evolution equations for force-free electrodynamics were developed, e.g., in \citet{Komissarov2004,McKinney2006}, and reviewed in further detail by \citet{Paschalidis2013}. Their conservation laws may be written in its vector form
\begin{align}
\partial_t\:\mathcal{C}+\partial_j\:\mathcal{F}^j=\left(\mathcal{S}_n+\mathcal{S}_s\right)\label{eq:ConsGeneral}\,,
\end{align}
where $\mathcal{C}$ denotes the conserved variables, $\mathcal{F}^j$ the flux vectors, $\mathcal{S}_n$ the geometrical and current induced source terms. Finally $\mathcal{S}_s$ are the potentially stiff \citep[see, e.g.][]{Leveque2002} source terms. Note that each of these quantities consists of elements in a multidimensional space. In general, the conserved variables $\mathcal{C}$ are derived from the so-called primitive variables.

Besides other strategies, the evolution of the full set of Maxwell equations of the fields $\left\{\mathbf{B},\mathbf{D},\mathbf{H},\mathbf{E}\right\}$ is introduced in the literature as a possibility to deal with electrodynamics in General Relativity by \citet{Komissarov2004}. In this case, an adaptation of eqs. (\ref{eq:MaxwellCovariantI}), and (\ref{eq:MaxwellCovariantII}) may be used as an evolutionary system. Following \citet{Palenzuela2009} as well as \citet{Mignone2010}, we suggest to study a suitably augmented system of Maxwell's equations:
\begin{align}
\nabla_\nu\left({^*}\hspace{-2pt}F^{\mu\nu}+s^{\mu\nu}\psi\right) =&\:-\kappa_\psi k^\mu\psi\label{eq:KomEvEqI}\\
\nabla_\nu\left(F^{\mu\nu}+g^{\mu\nu}\phi\right)=&\:I^\mu-\kappa_\phi k^\mu\phi\label{eq:KomEvEqII}
\end{align}
Here, we employ $s^{\mu\nu}=c_h^2\gamma^{\mu\nu}-n^\mu n^\nu$. The functions $\psi$ and $\phi$ are additional scalar potentials for the hyperbolic/parabolic cleaning of numerically induced divergence and charges, respectively \citep[cf.][]{Dedner2002,Palenzuela2009,Mignone2010}. $\kappa_{\psi}$, $c_h$, and $\kappa_{\phi}$ are the parameters controlling these cleaning terms. Contracting eq. (\ref{eq:KomEvEqI}) with $\nabla_\mu$ yields 
\begin{align}
c_h^2\nabla_i\nabla^i\psi+\nabla_t\nabla^t\psi=-\kappa_\psi\nabla_t\psi\,,
\end{align}
which compares to the telegraph equation. We stress the analogy of $c_h$ with a finite propagation speed for divergence errors \citep{Mignone2010} and their decay according to the damping factor $\kappa_\psi$. For $c_h$ equals to the speed of light, eq. (\ref{eq:KomEvEqI}) reduces to the evolution system given in \citet{Palenzuela2009}. The covariant Mawell equations (\ref{eq:KomEvEqI}), and (\ref{eq:KomEvEqII}) may be written in terms of the conserved quantities
\begin{align}
\arraycolsep=1.4pt\def\arraystretch{1.5}
\mathcal{C}\equiv\left(\begin{array}{c}
\mathcal{Q}\\ 
\mathcal{P}\\
b^i\\ 
d^i
\end{array}\right)=\: \sqrt{\gamma}\:\left(\begin{array}{c}
\frac{\psi}{\alpha}\\ 
\frac{\phi}{\alpha}\\
B^i+\frac{\psi}{\alpha}\beta^i\\ 
D^i-\frac{\phi}{\alpha}\beta^i
\end{array}\right)\label{eq:KCons}\,,
\end{align}
as well as the corresponding fluxes
\begin{align}
\arraycolsep=1.4pt\def\arraystretch{1.5}
\mathcal{F}^j\equiv\sqrt{\gamma}\:\left(\begin{array}{c}
B^j-\frac{\psi}{\alpha}\beta^j\\ 
-\left(D^j+\frac{\phi}{\alpha}\beta^j\right)\\
e^{ijk}E_k+\alpha\left(c_h^2\gamma^{ij}-n^i n^j\right)\psi\\ 
-\left(e^{ijk}H_k+\alpha g^{ij}\phi\right)
\end{array}\right)\label{eq:KFluxes}\,.
\end{align}
For the source terms, the split according to eq. (\ref{eq:ConsGeneral}) yields
\begin{align}
\arraycolsep=1.4pt\def\arraystretch{1.5}
\mathcal{S}_n\equiv\sqrt{\gamma}\left(\begin{array}{c}
\alpha\psi\Gamma^t_{\mu\nu}s^{\mu\nu}\\ 
-\alpha\phi\Gamma^t_{\mu\nu}g^{\mu\nu}-\rho\\
-\alpha\psi\Gamma^i_{\mu\nu}s^{\mu\nu}\\ 
-\alpha\phi\Gamma^i_{\mu\nu}g^{\mu\nu}- J^i
\end{array}\right)\:\:\text{and}\:\:
\arraycolsep=1.4pt\def\arraystretch{1.5}
\mathcal{S}_s\equiv\sqrt{\gamma}\left(\begin{array}{c}
-\alpha\kappa_\psi\psi\\ 
-\alpha\kappa_\phi\phi\\
0\\ 
0
\end{array}\right)\label{eq:KSources}\,.
\end{align}

In our practical implementation of the cleaning potential as potentially stiff source terms $\mathcal{S}_s$, we follow a Strang
splitting approach \citep[as employed, e.g. in][]{Komissarov2004}, effectively solving part of the scalar equations~(\ref{eq:KCons}) to (\ref{eq:KSources}) analytically. Prior (before \texttt{MoL\_Step}) and after (before \texttt{MoL\_PostStep}) the time integration of the Einstein Toolkit \texttt{thorn} \texttt{MoL} we evolve in time the equations
\begin{align}
\mathcal{P}\left(t\right)&=\mathcal{P}_0\exp\left[-\alpha^2\kappa_\phi t\right]\,,\\
\mathcal{Q}\left(t\right)&=\mathcal{Q}_0\exp\left[-\alpha^2\kappa_\psi t\right]\,,
\end{align}
for a time $t=\Delta t/2$. The coefficients $\kappa_\phi$ and $\kappa_\psi$ have to be chosen by optimisation in accordance with the grid properties. We find it beneficial to choose a large value for $\kappa_\phi$, effectively dissipating charge conservation errors on very short timescales. As for the divergence cleaning, we conducted a series of tests, optimizing $\kappa_\psi$ to yield stable and converging evolution for all shown resolutions, ultimately resorting to $\kappa_\psi = 0.125$. Numerical tests of the stability of force-free fields close to the BH horizon have shown that results improve significantly when advecting divergence cleaning errors $\psi$ faster than with $c_h=1$. In practice we, hence, employ $c_h=2$.

%--------------------------------------------------------------------
%--------------------------------------------------------------------
\subsection{The BZ process in 3D time-dependent GRFFE}
\label{sec:bz_signatures} 
%--------------------------------------------------------------------
%--------------------------------------------------------------------

In the following, we present a detailed study of the relations between energy flow, magnetic field strengths and field line angular velocity for the reference model (C-H4-L2). \citet{Blandford1977} quantify the energy extraction from (slowly) rotating BHs in \textit{stationary}, \textit{axisymmetric} force-free electrodynamics. The markers of BZ efficiency, like the field line angular velocity $\Omega_{\rm F}$ (\ref{eq:FL_AngVel}), and the toroidal magnetic field $B^T$ given by the Znajek condition (\ref{eq:ZnajekCondition}), do not need to be functions of the magnetic flux in 3D dynamical settings. The shown simulations accumulate points of inefficient or no energy extraction, where the ideal conditions described by \citet{Blandford1977} do not hold. Among the simulations that we present in this paper, we find many field lines for which the ideal conditions described by \citet{Blandford1977} do not hold and, hence, they are unable to efficiently extract energy out of the BH (or even extract energy at all). Figure~\ref{fig:field line_stats_znajek} points out combinations of reversed fields (with respect to the prescription in equation~\ref{eq:ZnajekCondition}) which have an energy inflow across the BH horizon rather than en energy outflow. Such field reversals in time-dependent 3D models can greatly reduce the overall process luminosity (\ref{eq:BZ_Luminosity}). In the results at hand, such a breakdown of process efficiency is naturally observed between the accretion of two consecutive loops, where the magnetospheric fields rapidly rearrange to change their polarity (see section \ref{sec:field_structure}, Figure~\ref{fig:counter_bpbr}). We find, indeed, that in our simulations an outgoing Poynting flux at the BH horizon positively correlates with the fulfillment of conditions derived by \citet{Blandford1977}, as we show in Figures~\ref{fig:pointcloud_poynting_velocity} and~\ref{fig:field line_stats_znajek}. In case of the counter-rotating disc models (series A and B, Table~\ref{tab:model_initials}) this outcome is especially noteworthy. Field lines depart with an angular velocity $\Omega_{\rm F}$ in the opposite direction to $\Omega_{\rm BH}$, but end up co-rotating under almost ideal conditions for the BZ process. For all models (co- and counter-rotating), we find the outflow of energy to be closely correlated to combinations of the magnetic fields $B^r$ and $B^\phi$ allowed by the Znajek condition (\ref{eq:ZnajekCondition}).

%--------------------------------------------------------------------
%--------------------------------------------------------------------
%\section{Blandford/Znajek signatures}
%\label{sec:bz_signatures} 
%--------------------------------------------------------------------
%--------------------------------------------------------------------

%--------------------------------------------------------------------
%--------------------------------------------------------------------
\subsection{Field line angular velocity}
\label{sec:field line_velocity}
%--------------------------------------------------------------------
%--------------------------------------------------------------------

\begin{figure}
	\centering
	\includegraphics[width=0.49\textwidth]{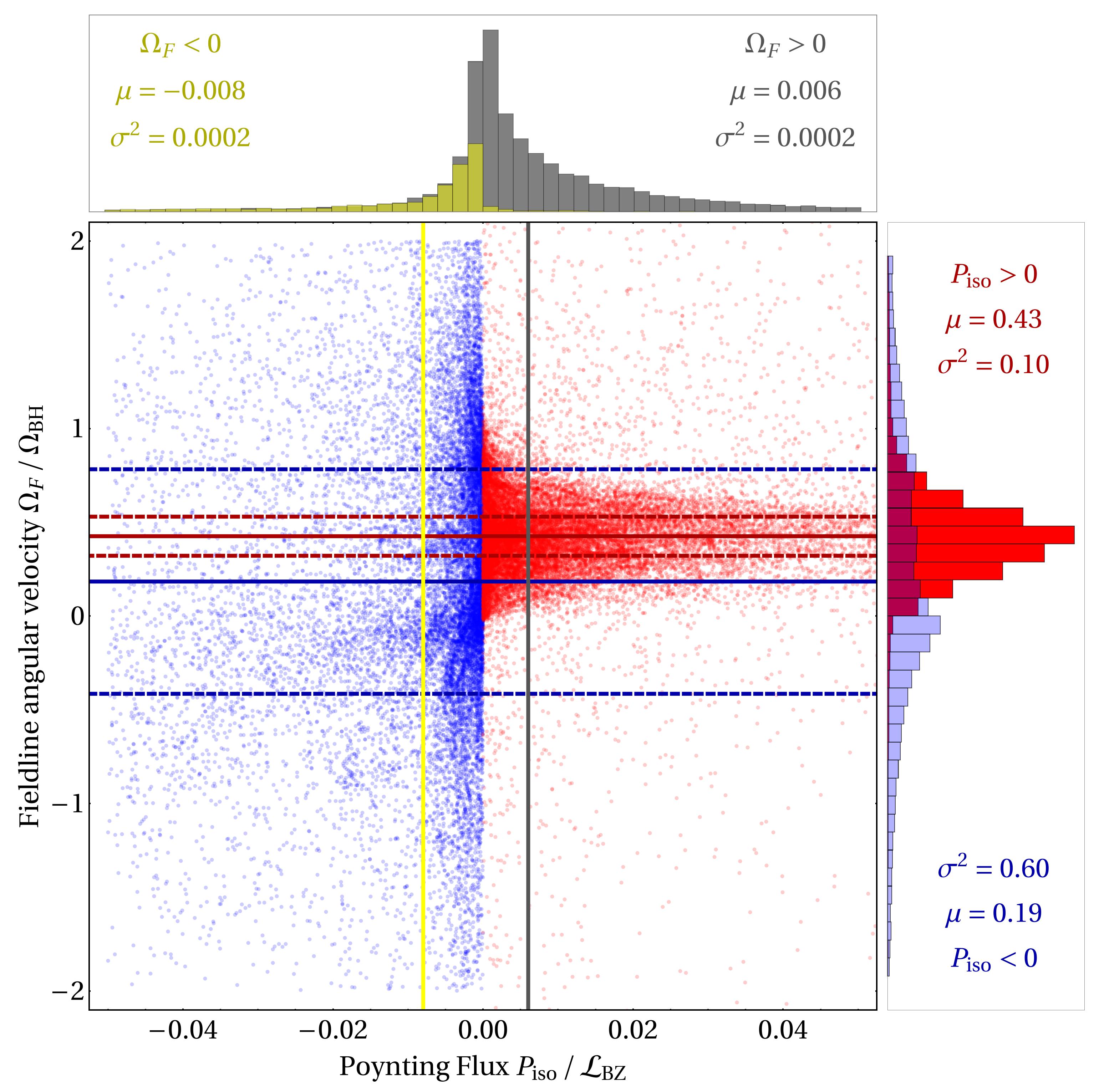}\hspace{5pt}
	\vspace{-15pt}
	\caption{Statistical analysis of the relationship between the (approximate) field line angular velocity $\Omega_{\rm F}/\Omega_{\rm BH}$ and  the isotropic Poynting flux (equation~\ref{eq:Piso}) normalized to the nominal BZ luminosity (equation~\ref{eq:BZ_Luminosity}) $P_{\rm iso}/\mathcal{L}_{\rm BZ}$. Sample points on the BH horizon have been extracted from the C-H4-L2-F model (414 time bins, 129 angular bins, 53.406 samples in total) and plotted individually in the graph. Basic statistical analysis is denoted by solid lines (mean $\mu$), and dashed lines (variance $\sigma^2$ at $\mu\pm \sigma$) if applicable. The outgoing Poynting flux is found mainly at field line velocities along the rotational sense of the BH and clustered around a mean value of $\Omega_{\rm F}/\Omega_{\rm BH}\approx 0.43$. The ingoing Poynting flux seems to be related (with large variance) to more negative field line angular velocities.}
	\label{fig:pointcloud_poynting_velocity}
\end{figure}
\begin{figure}
	\centering
	\includegraphics[width=0.425\textwidth]{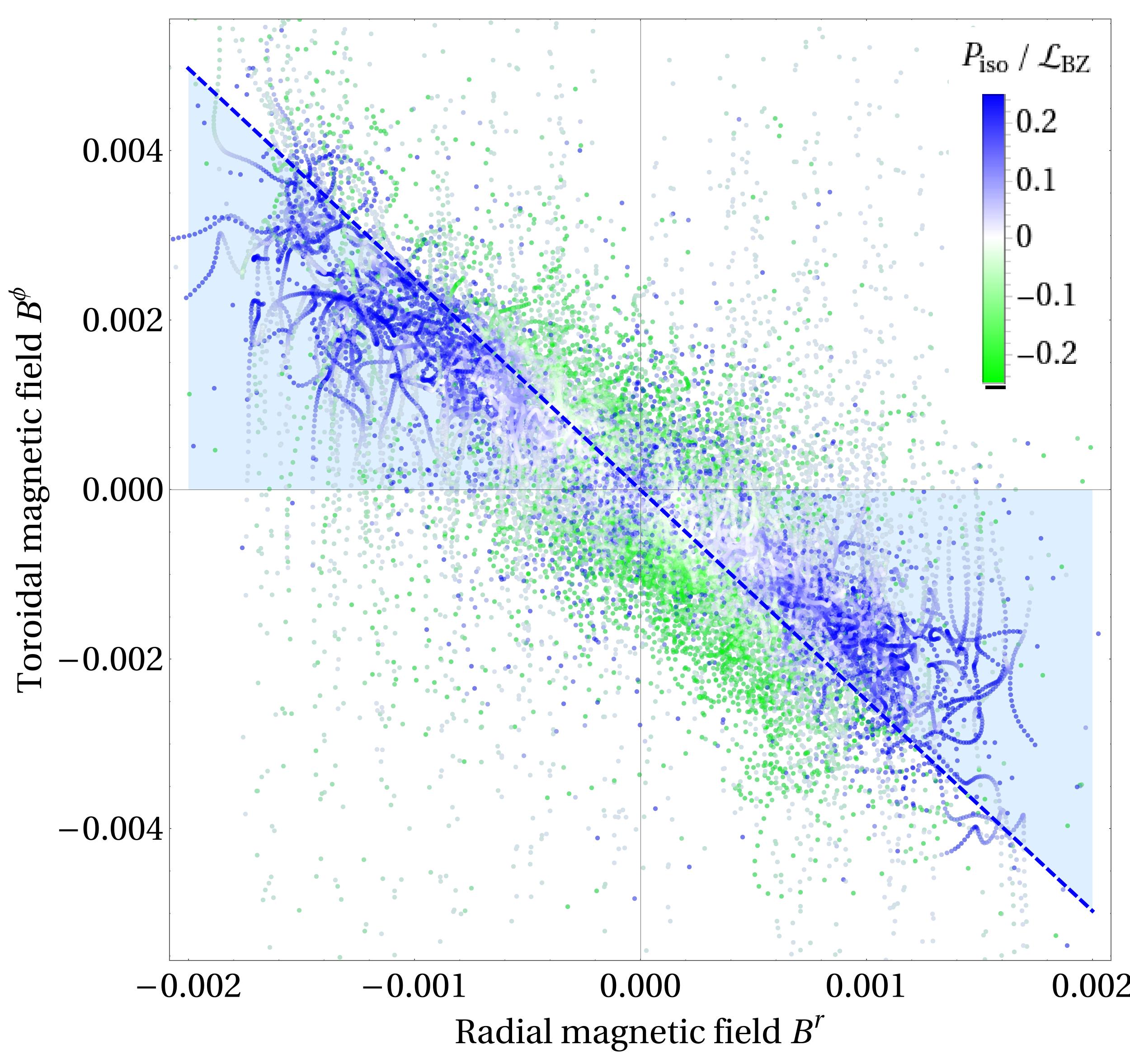}\hspace{5pt}
	\vspace{-4pt}
	\caption{Statistical analysis of the relationship between the radial and toroidal magnetic fields $B^r$ and $B^\phi$. Sample points as in Figure~\ref{fig:pointcloud_poynting_velocity}. The color-scale indicates the respective ingoing (\textit{green}) or outgoing (\textit{blue}) Poynting flux. The combinations allowed by the Znajek condition (equation~\ref{eq:ZnajekCondition}) for a field line angular velocity co-rotating with the BH (see figure~\ref{fig:pointcloud_poynting_velocity}) is denoted by the \textit{blue dashed} line and the corresponding shading. An outgoing Poynting flux seems to be strongly correlated with a fulfillment of the Znajek condition, while an ingoing Poynting flux is arbitrarily scattered. The latter may occur especially for a fields which are reversed in sign with respect to condition (\ref{eq:ZnajekCondition}).}
	\label{fig:field line_stats_znajek}
\end{figure}

The ratio between the angular velocity of the central object $\Omega_{\rm BH}$ and magnetic field lines threading the BH horizon $\Omega_{\rm F}$ is key for an efficient Poynting flux dominated energy extraction \citep[see equation~4.5 in][]{Blandford1977}. More specifically, $\Omega_{\rm F}=\Omega_{\rm BH}/2$ is the optimal value and was used in the derivation of equation~(\ref{eq:BZ_Luminosity}). In practice, studying more realistic (e.g., paraboloidally shaped) stationary and axisymmetric field lines in force-free Kerr magnetospheres, \citet{Nathanail2014} and \citet{Mahlmann2018} find $\Omega_{\rm F}$ to be smaller than this ideal value ($\sim 0.45\Omega_{\rm BH}$) for BHs with $M=1$ and $a=0.9$. In fact, \citet{Blandford1977} examine a paraboloidal field line configuration with $\Omega_{\rm F}\in\left[0.27\Omega_{\rm BH},0.5\Omega_{\rm BH}\right]$. In \textit{stationary}, \textit{axisymmetric} force-free electrodynamics, a field line angular velocity $\Omega_{\rm F}$ may be defined employing the relations given in section~\ref{sec:force-free} \citep{Blandford1977,Carrasco2017}:
\begin{align}
\Omega_{\rm F}=\frac{F_{t\theta}}{F_{\theta\phi}}=-\frac{E_{\theta}}{\sqrt{\gamma}B^r}\label{eq:FL_AngVel}
\end{align}
As the presented numerical simulations are \textit{fully dynamic} and \textit{3D}, equation~(\ref{eq:FL_AngVel}) will only have limited applicability. However, it was used, e.g. by \citet{Yuan2019} in order to estimate $\Omega_{\rm F}$ during axisymmetric and relaxed episodes of BH force-free magnetospheric evolution. In the following, we will employ some basic statistical analysis in order to make statements about the correlation of field line angular velocity and energy extraction. Figure~\ref{fig:pointcloud_poynting_velocity} displays a sample of combinations of an isotropic power emerging from an angular patch $\theta\in\left[\theta_-\:,\:\theta_+\right]$,
\begin{align}
P_{\rm iso}\left(\theta\right)=\frac{2\pi}{\cos\theta_+-\cos\theta_-}\int_{\theta_-}^{\theta_+}S_+^r\,\text{d}A_r,
\label{eq:Piso}
\end{align}
and $\Omega_{\rm F}$ at different azimuthal positions and times throughout one of the conducted simulations. We observe clear trends in their correlation, for example examining model C-H4-L2 (see figure~\ref{fig:pointcloud_poynting_velocity}):
\begin{enumerate}
	\item Positive field line angular velocities ($\Omega_{\rm F}>0$) on average correspond to outgoing energy flux. Conversely, $\Omega_{\rm F}<0$ typically correlate with an ingoing, weaker Poynting flux. 
	\item The extraction of power ($P_{\rm iso}>0$) is clearly clustered around $\Omega_{\rm F}\approx 0.43\Omega_{\rm BH}$, while its ingoing counterpart ($P_{\rm iso}<0$) corresponds to mean values $\Omega_{\rm F}\approx 0.19\Omega_{\rm BH}$ and has a sixfold greater variance.
\end{enumerate}

Qualitatively similar statistical results hold for all counter-rotating models (series A and B, Table~\ref{tab:model_initials}). In the case of co-rotating ADs, however, there is no clear correlation between positive angular velocities and outgoing Poynting flux. The average field line angular velocity at the BH horizon for these models (series C) usually is $\Omega_{\rm F}>\Omega_{\rm BH}/2$, across both inward and outward flowing Poynting flux.

%--------------------------------------------------------------------
%--------------------------------------------------------------------
\subsection{Znajek condition}
\label{sec:znajek_condition}
%--------------------------------------------------------------------
%--------------------------------------------------------------------

The so-called Znajek condition  \citep{Blandford1977,Znajek1977} must be satisfied by the magnetic field at the BH horizon to enable a positive energy extraction in the BZ mechanism. The existence of asymptotic conditions for magnetic fields in stationary MHD was first posed by \citet{Weber1967} and applied to Kerr BHs by \citet{Znajek1977}. In order to ensure finite field quantities at the BH horizon, the Znajek condition requires the following relationship between the radial and toroidal magnetic fields $B^r$ and $B^\phi$:
\begin{align}
B^\phi=\frac{\sin\theta}{\alpha}\frac{\Omega_{\rm F}\left(r_+^2+a^2\right)-a^2}{r_+^2+a^2\cos^2\theta}B^r
\label{eq:ZnajekCondition}
\end{align}
While \cite{Blandford1977} employ condition~(\ref{eq:ZnajekCondition}) as a boundary condition at the BH horizon, it is nowadays understood as a regularity condition at the BH horizon \citep[e.g.][]{Komissarov2004,okamoto2006,Beskin2010,Contopoulos2013}. As such it is formulated in \textit{stationary}, \textit{axisymmetric} force-free electrodynamics, i.e. there is no intrinsic guarantee for its fulfillment in 3D time-dependent GRFFE. Figure~\ref{fig:field line_stats_znajek} shows that the presented simulations assemble an unrestricted range of possible combinations between $B^r$ and $B^\phi$. However, we find (for all employed models) that an outgoing Poynting flux is favored by combinations allowed by the Znajek condition (\ref{eq:ZnajekCondition}) and stronger overall magnetic fields.

%--------------------------------------------------------------------
%--------------------------------------------------------------------
\section{Benchmark to similar research}
\label{sec:yuan_benchmark} 
%--------------------------------------------------------------------
%--------------------------------------------------------------------

\begin{figure*}
	\centering
	\includegraphics[width=0.98\textwidth]{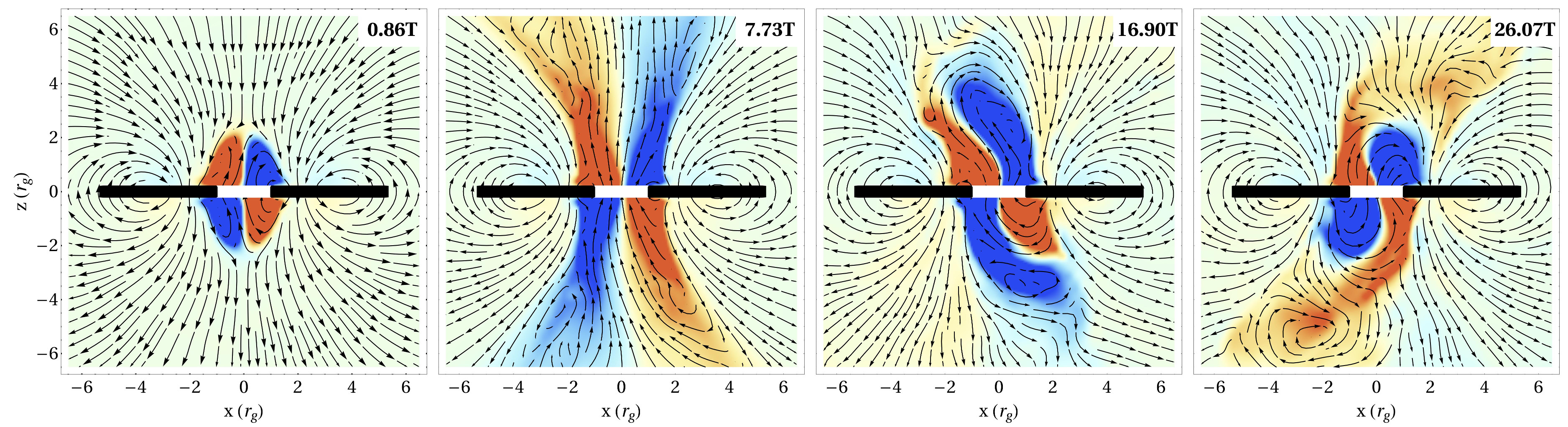} 
	\vspace{-6pt}
	\caption{Snapshots from toy-models employed to compare our results to those of  \citet{Yuan2019a}. The models consist of disc-shaped membranes (located at $z=0$) emulating both the central object (a BH or generically a star) surrounded by a thin disc. The colours represent the toroidal field $B^\phi$ component, which is perpendicular to the displayed $x-z$ plane (\textit{blue} negative, \textit{red} positive). Streamlines display the poloidal magnetic field (components in the $x-z$ plane). The central membrane (thick white line) corresponds to a perfectly conducting object, $\eta=0$, with cylindrical radius $r_1=1$, angular velocity $r_1\Omega_{\rm BH}=0.9$. The outer membrane (thick black line) is a truncated disc, that is perfectly conducting ($\eta=0$), non rotating ($\Omega_{\rm D}=0$), and extends outward from an inner radius $r_2=1.08r_g$ with decay parameter $\alpha=0.4$ (see equation~\ref{eq:currentYuan}). The dipolar-shaped magnetic field of the disc is driven by a current loop on the equatorial plane located at a distance $r_0=4.33r_g$. Field lines connecting the central membrane to the disc are twisted up showing kink structures and stripes of reversed fields. The panels denote the simulation time after initialization in units of the rotational period $\text{T}=2\pi/\Omega_{\rm BH}$. The grid resolution on the finest refinement level is $\Delta_{x,y,z}= 0.125$. The color scale is the same over all plots.} 
	\label{fig:rigid_rotator}
\end{figure*}
\begin{figure*}
	\centering
	\includegraphics[width=0.98\textwidth]{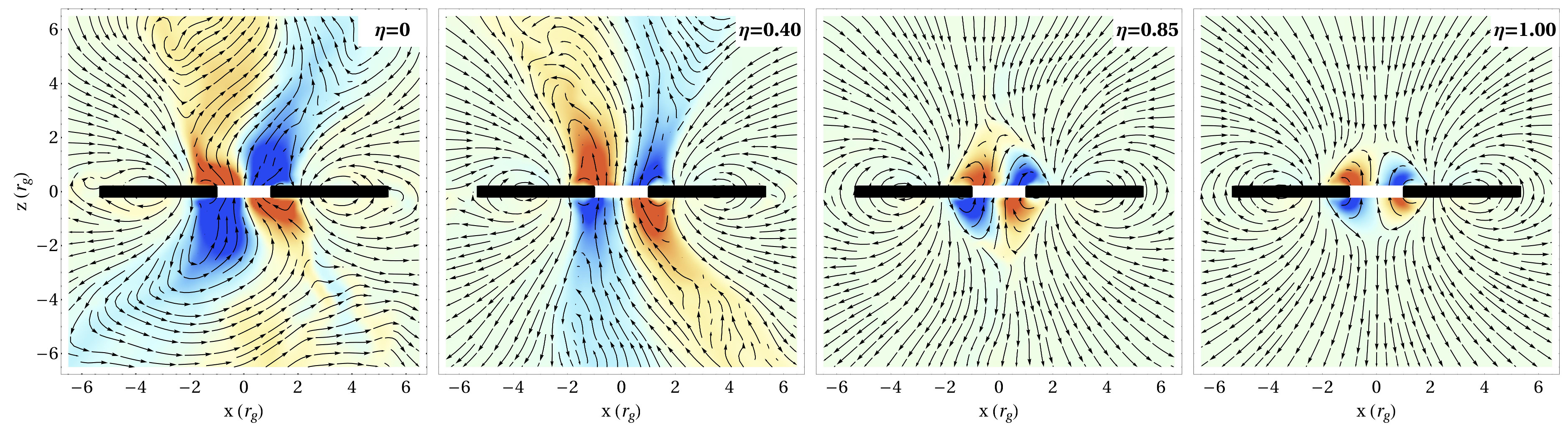} 
	\caption{As in Figure~\ref{fig:rigid_rotator} but with decay parameter $\alpha=0.9$ and varying surface resistivity $\eta$ of the central membrane. Snapshots are shown for $t=26.07\text{T}$.}
	\label{fig:membrane_rotator}
\end{figure*}

\citet{Yuan2019} present a model to study 3D Minkowski dynamical force-free magnetospheres with field lines anchored to a central object and an extended AD. In order to include the electrodynamical effects of a BH on the AD, \cite{Yuan2019} resort to the membrane paradigm to build a BH/AD analog, where both the BH and the AD are replaced by thin, equatorial membranes. In their setup, \citet{Yuan2019} assume that the BH is rotating with an angular velocity $\Omega_{\rm BH}$, while the AD may spin with an angular velocity $\Omega_{\rm D}$. Thus, field lines connecting the BH and the AD are differentially rotating. Prescribing the surface resistivity $R$ of a disc-shaped central membrane (representing the BH) of radius $r_1$ and angular velocity $\Omega_{\rm BH}$, they are able to mimic different degrees of co-rotation of field lines of the BH/AD system. Following the membrane paradigm, the BH horizon has a surface resistivity of $R=4\pi/c\text{ (Gaussian)}=1\text{ (geom.)}\approx 377\Omega\text{ (SI)}$ \citep{Thorne1986}. The AD is set up as a perfect conductor with a surface resistivity $R=0$. The field lines are dragged along rigidly with the AD angular velocity $\Omega_{\rm D}$. A structure of two tubes of zero-net magnetic flux is initialized by the following surface current in the (equatorial) AD:
\begin{align}
\sqrt{g_{\phi\phi}}\,J^\phi=\hspace{-0.1cm}\left\{\begin{array}{cc}
 \hspace{-0.15cm} J_0 \cos\left(\displaystyle\frac{2\pi}{r_0}\left(r-r_2\right)\right)\left(\displaystyle\frac{r_2}{r}\right)^\alpha, & r_2\leq r\leq r_2+\displaystyle\frac{3}{4}r_0 \\ 
0, & \text{else}
\end{array}\right.
\label{eq:currentYuan}
\end{align}
where $r_2$ is the inner radius of the equatorial disc, $r_0$ is the center of the the current loop that generates a dipolar-shaped magnetic field in the disc, and the exponent $\alpha$ controls de radial decay of the boundary. $J_0$ is a normalization constant, which we take equal to 0.1.
This current is similar to the one employed in the presented simulations (cf. equation~\ref{eq:InitCurrent}), differing mainly by a shift along the equatorial direction. After an initialization period, the resulting magnetic field is fixed as a boundary condition along the AD and the initialization current is replaced by suitable restrictions on the electric field \citep[see][for further details]{Yuan2019}. In practice, the two membranes do not have zero thickness, but extend over a few cells of the numerical grid. We reset the space charge $\rho$ and the numerical cleaning potentials to zero across all the numerical cells encompassing the membranes. We reproduce two exemplary series of setups from \citet{Yuan2019} for a benchmark with the presented GRFFE method (Figures~\ref{fig:rigid_rotator} and~\ref{fig:membrane_rotator}). Our results qualitatively reproduce those of \cite{Yuan2019}. A quantitative comparison is not possible since these authors do not provide enough information on the numerical grid and on the constants employed to setup their models. For a closer comparison with the BH/AD model described in section~\ref{sec:loop_systems}, we repeat a similar test with appropriate changes to the setup in section~\ref{sec:discussion} (see Figure~\ref{fig:strength_model}). There, due to the full GR capacities of our method, we do not set further boundaries on the central BH (as our coordinates are horizon penetrating). Field lines rotate rigidly with the AD by enforcing zero space charge and the force-free electric field \citep[cf, e.g.,][]{Camenzind2007}
\begin{align}
D^r&=\frac{\sqrt{\gamma}}{\alpha}\gamma^{rr}\left(\Omega_{\rm D}+\beta^\phi\right)B^\theta,\\
D^\theta&=-\frac{\sqrt{\gamma}}{\alpha}\gamma^{\theta\theta}\left(\Omega_{\rm D}+\beta^\phi\right)B^r,\\
D^\phi&=0\vphantom{-\frac{\sqrt{\gamma}}{\alpha}\gamma^{\theta\theta}\left(\Omega_{\rm D}+\beta^\phi\right)}.
\end{align}
%

%--------------------------------------------------------------------
%--------------------------------------------------------------------
\section{Energy outflow from the disc}
\label{sec:faraday_disc}
%--------------------------------------------------------------------
%--------------------------------------------------------------------

We find that our simplified AD model also artificially generates a radially outwards directed Poynting flux. The luminosity of the disc is very different in prograde and retrograde discs, being negligible for the latter cases, but significant for the former ones. An estimate of the flow of energy emerging from an isolated AD can be found in the so-called Faraday disc \citep[e.g.,][and references therein]{Feynman2011,Chyba2015}. A disc-shaped electrical generator of height $h_{\rm F}$ rotating with angular velocity $\omega_{\rm F}$ in a uniform magnetic field $B_{\rm F}$ relies on the energy conversion from the magnetic field to the induced electric current $I_{\rm F}$ by the mechanical rotation. On the other hand, the current $I_{\rm F}$ brakes down the rotation the of the Faraday disc ensuring energy conservation. In our case we impose \textit{both} disc rotation $\omega_{\rm F}$, and the disc current $I_{\rm F}$ in the \textit{ad hoc} disc model (Figure~\ref{fig:discsetup}). Energy conservation, i.e. the interplay between induced electric current and magnetic braking, demands that energy $\mathcal{E}_{\rm F}$ be radiated from the disc. The amount of energy radiated can be roughly estimated as:
	\begin{align}
	\mathcal{E}_{\rm F}=\frac{\pi}{2}\sigma B_{\rm F}^2\:\omega_{\rm F}^2\:h_{\rm F}\:\left(b^4-a^4\right)
	\label{eq:Faraday_Disk}
	\end{align}
Here, $a$ and $b$ represent the inner and outer radius of the Faraday disc, respectively. For magnetic fields $B_{\rm F}^2\propto 1/r^4$ and $\omega_{\rm F}^2\propto 1/r^{3}$ (cf. equation~\ref{eq:VelocityOuter}) the energy radiated by a disc of finite length $|b-a|$ - corresponding, e.g. to $\sim l$ in one flux tube in our simulations - scales as $1/r^3$. Apart from the exact field geometry in the AD model, the power induced into the magnetosphere by this artificial process (equation~\ref{eq:Faraday_Disk}) is greatly suppressed for counter-rotating discs due to the remote location of $r_{\rm ISCO}$ from the central object. Conversely, for prograde discs, in which $a\simeq r_{\rm ISCO}\sim r_+$, the AD luminosity of our models may even be larger than BZ luminosity. This is the case for large enough BH rapidity ($a^*\gtrsim 0.9$).

\label{lastpage}

\end{document}